\documentclass[prx,epsf,showpacs,twocolumn,superscriptaddress,footinbib]{revtex4-2}

\usepackage[pdftex]{graphicx}
\usepackage{dcolumn}
\usepackage{bm}
\usepackage{comment}
\usepackage{epsfig}
\usepackage{latexsym}
\usepackage{amsmath}
\usepackage{amssymb}
\usepackage{amsfonts}
\usepackage[position=top,caption=false,singlelinecheck=off, justification=raggedright]{subfig}
\usepackage{color}
\usepackage{array}
\usepackage{braket}
\usepackage{bbold}
\usepackage{dsfont}
\usepackage{float}
\usepackage{mathrsfs}
\usepackage{xcolor} % For defining custom colors
\usepackage{listings} 
\usepackage{tikz}
\usepackage[inline]{enumitem}
\usetikzlibrary{automata, positioning, arrows, quantikz}
\usepackage{hyperref}
\hypersetup{
    colorlinks,
    citecolor=blue,
    filecolor=red,
    linkcolor=magenta,
    urlcolor=blue
}
\newcommand{\eps}{\epsilon}

\makeatletter
\DeclareFontFamily{OMX}{MnSymbolE}{}
\DeclareSymbolFont{MnLargeSymbols}{OMX}{MnSymbolE}{m}{n}
\SetSymbolFont{MnLargeSymbols}{bold}{OMX}{MnSymbolE}{b}{n}
\DeclareFontShape{OMX}{MnSymbolE}{m}{n}{
    <-6>  MnSymbolE5
   <6-7>  MnSymbolE6
   <7-8>  MnSymbolE7
   <8-9>  MnSymbolE8
   <9-10> MnSymbolE9
  <10-12> MnSymbolE10
  <12->   MnSymbolE12
}{}
\DeclareFontShape{OMX}{MnSymbolE}{b}{n}{
    <-6>  MnSymbolE-Bold5
   <6-7>  MnSymbolE-Bold6
   <7-8>  MnSymbolE-Bold7
   <8-9>  MnSymbolE-Bold8
   <9-10> MnSymbolE-Bold9
  <10-12> MnSymbolE-Bold10
  <12->   MnSymbolE-Bold12
}{}

\let\llangle\@undefined
\let\rrangle\@undefined
\DeclareMathDelimiter{\llangle}{\mathopen}%
                     {MnLargeSymbols}{'164}{MnLargeSymbols}{'164}
\DeclareMathDelimiter{\rrangle}{\mathclose}%
                     {MnLargeSymbols}{'171}{MnLargeSymbols}{'171}

\newcommand{\beginappendix}{%
        \setcounter{table}{0}
        \renewcommand{\thetable}{A\arabic{table}}%
        \renewcommand{\theHtable}{Appendix\thetable}
        \setcounter{equation}{0}
        \renewcommand{\theequation}{A\arabic{equation}}%
        \renewcommand{\theHequation}{Appendix\theequation}
        \setcounter{figure}{0}
        \renewcommand{\thefigure}{A\arabic{figure}}%
        \renewcommand{\theHfigure}{A\thefigure}
        \setcounter{section}{0}
        \renewcommand{\thesection}{A\arabic{section}}%
     }

\definecolor{RED}{rgb}{1,0,0}

\definecolor{BLUE}{rgb}{0,0,1}

\definecolor{codegreen}{rgb}{0,0.6,0}
\definecolor{codeblue}{rgb}{0,0,0.6}
\definecolor{codegray}{rgb}{0.5,0.5,0.5}
\definecolor{codepurple}{rgb}{0.58,0,0.82}
\definecolor{backcolour}{rgb}{0.95,0.95,0.92}

\usepackage[normalem]{ulem}

\begin{document}

\title{Helios: A 98-qubit trapped-ion quantum computer}

\def\cambridge{Quantinuum, Cambridge, CB2 1NL, UK}
\def\london{Quantinuum, London SW1P 1BX, UK}
\newcommand{\QTN}{Quantinuum, Broomfield, CO 80021, USA}
\newcommand{\QTNBP}{Quantinuum, Brooklyn Park, MN 55422, USA}
\newcommand{\QTNPM}{Quantinuum, Plymouth, MN 55422, USA}
\newcommand{\QTNKK}{Quantinuum K.K., Tokyo, Japan}
\newcommand{\QTNUK}{ Quantinuum, London, UK}
\newcommand{\QTNC}{ Quantinuum, Cambridge, UK}
\newcommand{\SandiaCA}{Quantum Performance Laboratory, Sandia National Laboratories, Livermore, CA 94550}
\newcommand{\SandiaNM}{Quantum Performance Laboratory, Sandia National Laboratories, Albuquerque, NM 87185}

% % for first posting on website
% \author{Quantinuum}
% \affiliation{\QTN}
% \affiliation{\QTNBP}
% \affiliation{\QTNPM}
% \affiliation{\QTNKK}
% \affiliation{\QTNUK}
% \affiliation{\QTNC}
% \author{Quantum Performance Laboratory, Sandia National Laboratories}
% \affiliation{\SandiaCA}
% \affiliation{\SandiaNM}

% --- Lead authors (order preserved) ---
\author{Anthony Ransford}
\email{anthony.ransford@quantinuum.com}
\affiliation{\QTN}
\author{M.S. Allman}
\affiliation{\QTN}
\author{Jake Arkinstall}
\affiliation{\QTNC}
\author{J.P. Campora III}
\affiliation{\QTN}
\author{Samuel F. Cooper}
\affiliation{\QTN}
\author{Robert D. Delaney}
\affiliation{\QTN}
\author{Joan M. Dreiling}
\affiliation{\QTN}
\author{Brian Estey}
\affiliation{\QTN}
\author{Caroline Figgatt}
\affiliation{\QTN}
\author{Alex Hall}
\affiliation{\QTN}
% ...

\begin{abstract}
We report on Quantinuum Helios, a 98-qubit trapped-ion quantum processor based on the quantum charge-coupled device (QCCD) architecture. Helios features $^{137}$Ba$^{+}$ hyperfine qubits, all-to-all connectivity enabled by a rotatable ion storage ring connecting two quantum operation regions by a junction, speed improvements from parallelized operations, and a new software stack with real-time compilation of dynamic programs. Averaged over all operational zones in the system, we achieve average infidelities of $2.5(1)\times10^{-5}$ for single-qubit gates, $7.9(2)\times10^{-4}$ for two-qubit gates, and $4.8(6)\times10^{-4}$ for state preparation and measurement, none of which are fundamentally limited and likely able to be improved. These component infidelities are predictive of system-level performance in both random Clifford circuits and random circuit sampling, the latter demonstrating that Helios operates well beyond the reach of classical simulation and establishes a new frontier of fidelity and complexity for quantum computers.

\vspace{1.5in}

\end{abstract}

\maketitle

\section{Introduction}

Quantum computing hardware has progressed significantly in the last decade, providing strong experimental evidence of quantum supremacy~\cite{arute2019,PhysRevLett.127.180501, DeCross2025} and the feasibility of fault-tolerance~\cite{RyanAnderson2021,Acharya2025}. As an increasing number of different modalities check off the basic requirements for quantum computing, the focus of progress is shifting toward scaling these systems to much larger sizes without sacrificing performance.

Like all modalities, the trapped-ion QCCD architecture~\cite{Wineland98,Kielpinski02,Home2009,Pino2020,moses2023race, Mordini2025} has a unique set of engineering challenges in scaling. For example, trapped-ions can require laser systems for loading, cooling, state-preparation, measurement and coherent control (or a subset of these), introducing somewhat non-standard integration constraints between sub-systems. However, atomic-qubit architectures that use qubit transport, including QCCD and optical tweezers~\cite{Bluvstein2022, reichardt2025faulttolerantquantumcomputationneutral}, can distribute these computational resources more efficiently than stationary qubits. Mobile qubit architectures allow qubits to flow through the QPU like bits in classical processing architectures, with separated memory structures, data buses, and logic processing units, each optimized for their function. Conversely, stationary-qubit architectures, like superconducting qubits~\cite{arute2019, IBM_utility} or even atomic-qubits without transport~\cite{InfleqtionArch, Chen2024benchmarkingtrapped}, deliver quantum operations to each individual qubit (or connected qubits), which can pose significant engineering and calibration issues. Since transport-based qubits can share expensive hardware resources, the relative complexity of optical control systems (for example) is largely offset by reducing the multiplicative complexity associated with the number of processing zones \cite{Brandl2017}. Of course, the effectiveness of mobile qubit design principles depends on how sensitive the quantum information is to the required transport operations and how easy the controls are to build and operate. Using hyperfine clock-states and standard scalable micro-fabricated traps for transport control, the QCCD architecture can readily take advantage of these strategies. Indeed, as we show in this work, trapped-ion QPUs are roughly scaling in qubit number as fast or faster than solid state technologies, with the first QCCD computer demonstrated five years ago with 6 qubits~\cite{Pino2020}, to now using 98 qubits.

\begin{figure}[t]
    \centering
    \includegraphics[width=1\linewidth]{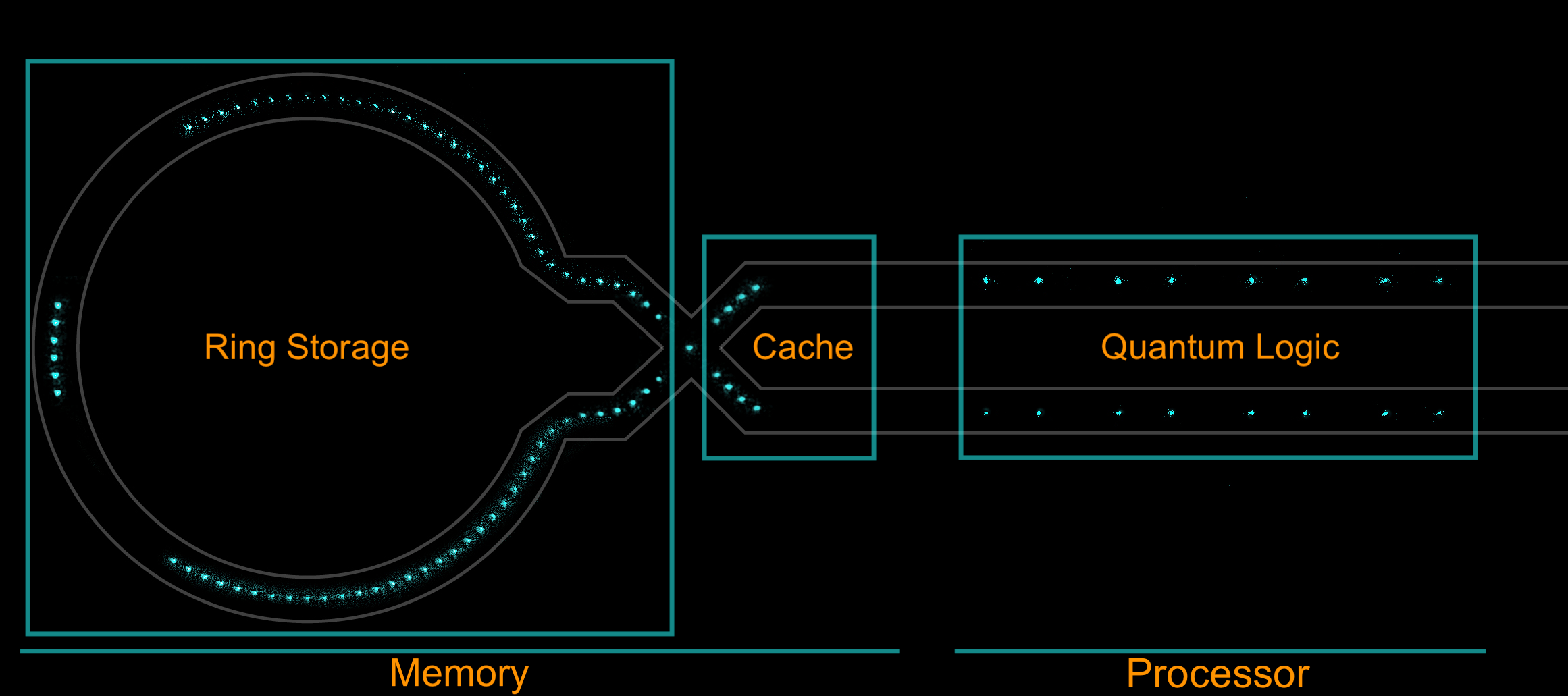}    
    \caption{An image of 98 atomic ions illuminated by resonant laser light in the Helios 2D surface trap illustrated in Fig.~\ref{fig:MainFig}. The overlaid lines indicate different regions of the device with the quincunx of ions showing the location of the ion trap junction.} 
    \label{fig:ions_image}
\end{figure}

\begin{figure*}[t]
    \centering
    \includegraphics[width=1\linewidth]{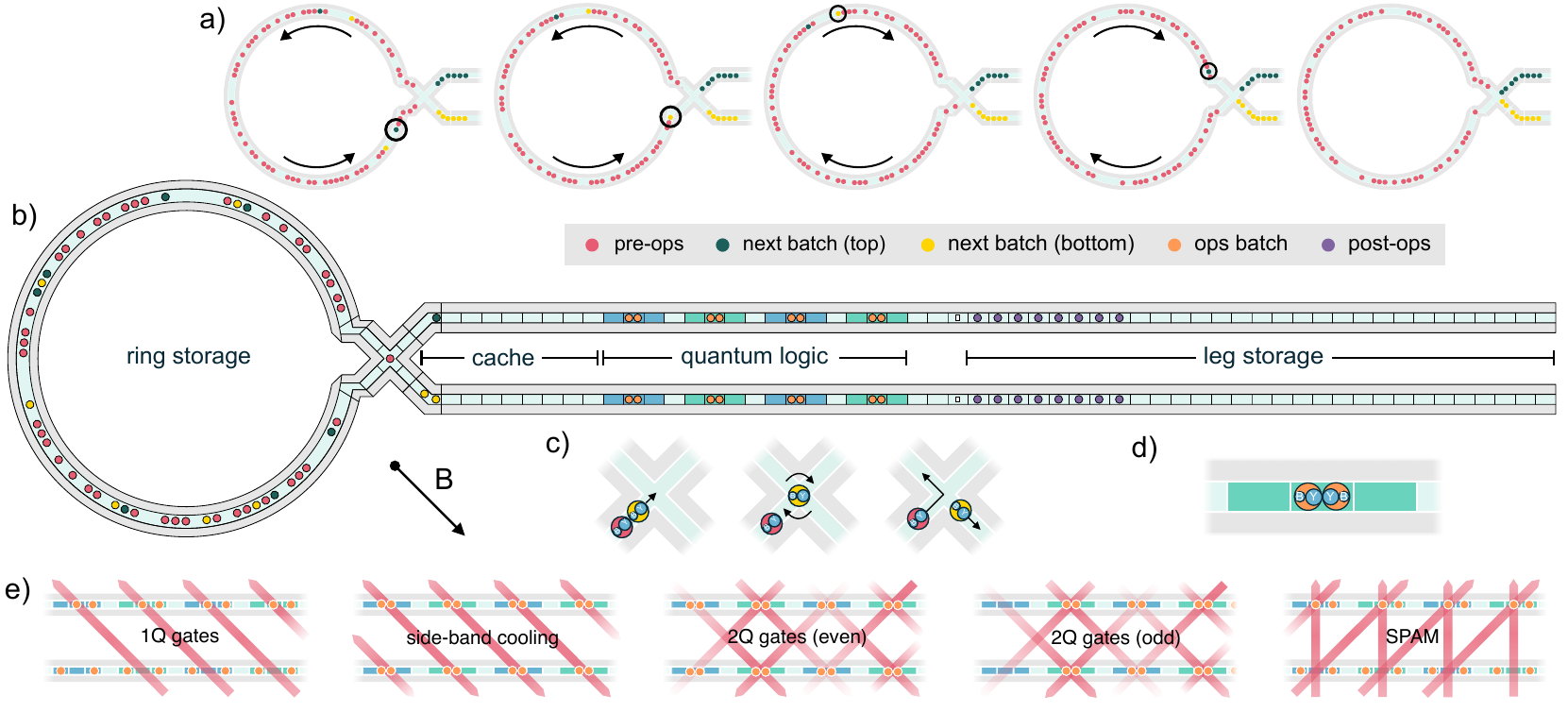}
    \caption{An illustration of the Helios design and conception of operations. (a) The final five stages of loading the cache region with qubits from ring storage. The ring rotates ions in both directions to move the circled qubit into the cache. (b) Diagram of trap (not to scale) part-way through a program, ring storage qubits are being loaded into the cache and qubits in the quantum logic region are undergoing ground-state cooling. The actual horizontal length is 15.3 mm, the ring diameter is 2.8 mm, and the operational zones are 750 $\mu$m apart. (c) Junction operations showing the retrieval and alignment of an ion crystal, and an ion crystal moving through the junction to stay in the ring storage. (d) The proper alignment of a 4-ion crystal in the quantum logic zones. (e) Laser beam and crystal configurations during example quantum operations as labeled. Beams are focused to operate on top/bottom legs as shown by color gradients. The 2Q gate beams are tilted both vertically and horizontally away from the 45 degree line that intersects the ion crystals in both legs by approximately 1 degree so to only interact with a single ion crystal at a time.}
    \label{fig:MainFig}
\end{figure*}

In this paper, we present Helios, the next generation system from Quantinuum, which introduces three advances to transport-based, trapped-ion quantum computers. First, Helios uses barium ions as the qubits \cite{dietrich2010hyperfine}, achieving improved quantum operation error rates with a more scalable laser architecture compared to ytterbium ions used in earlier Quantinuum QPUs~\cite{Pino2020, moses2023race}. Second, we use a four-way ``X" junction~\cite{Burton2023Junction, Hensinger2006,Blakestad2009,Decaroli_2021,Blakestad2011,Wright_2013,Amini_2010,Moehring_2011,Shu2014} to efficiently connect memory regions to quantum logic regions without increasing electrical control or device fabrication complexity compared to the Quantinuum H2~\cite{moses2023race}. Third, Helios is orchestrated by a new classical control implementation capable of making real-time decisions about all transport and quantum operations, enabling execution of truly arbitrary quantum programs with all-to-all connectivity. We show that these generational advancements set a new state-of-the-art in digital quantum computers according to several figures of merit---average two-qubit, single-qubit, and state preparation and measurement (SPAM) fidelities---confirmed by component and system-level benchmarks.

We organize this paper by first providing an overview of the notable advances in the Helios system---the architecture and trap design \ref{sec:architecture}, the ion species \ref{sec:ba_qubit}, the concept of operations of the QCCD \ref{sec:TransportAndParallelOps} and the real time compilation of programs \ref{sec:compiler}. We then establish the performance of the QPU with component-level and system-level benchmarking data in Sec.~\ref{sec:benchmarking}.  Finally, we discuss the outlook in Sec.~\ref{sec:outlook} and provide additional experimental and theoretical details in the appendix.

\section{Hardware and Software Architecture} \label{sec:h2}

\subsection{QPU architecture and ion trap design } \label{sec:architecture} \vspace{0.5em}

Helios is a transport-based quantum processor with spatially separated qubit memory regions and quantum logic regions. These elements are realized on a 2D surface electrode QCCD~\cite{Chiaverini2005,Pino2020}, which confines ions with electric fields generated by a pattern of electrodes (see Fig.\,\ref{fig:ions_image} and Fig.\,\ref{fig:MainFig}), and the QPU uses individual ions for qubits. To apply gates to qubits or pairs of qubits, the ions are physically transported to isolated trapping zones to facilitate low-crosstalk addressing and maintain high fidelity.

Figure \ref{fig:MainFig} illustrates how Helios operates. The quantum logic region processes batches of up to 16 qubits at a time, using 8 high-fidelity operation zones, each with the capability to perform state preparation, measurement, ground-state laser cooling, and quantum logic gates. Each operation is implemented via focused laser beams propagating parallel to the chip surface as shown in Fig.~\ref{fig:MainFig}e. High-fidelity operation necessitates low noise, independent electrode voltages and multiple laser beams for each zone, so they consume most of the control resources in the processor. By using shared lasers across multiple operation zones (Fig.~\ref{fig:MainFig}e), the quantum logic region design scales these essential components more efficiently than previous systems.

Qubits outside the operation zones are stored in functionally distinct memory regions:  ring storage, leg storage, and cache, see Fig.~\ref{fig:MainFig}b. Memory regions require less control resources as the only operations available are sympathetic laser cooling~\cite{Kielpinski2000} and qubit transport. To minimize the number of transport control signals, segmented DC electrodes in the memory regions use voltages that are broadcast in a repeating triplet pattern similar to Ref.~\cite{moses2023race}. The cache is a small memory region that holds the next batch of pre-sorted qubits before going to the quantum logic region. The leg storage operates as a first in, last out memory, while the ring storage acts as a random access memory, because it connects to the operational region via an X-junction.

The junction is a key structure enabling this architecture. As qubits move through the junction, they can be routed to remain in memory or be added to the cache in either the upper or lower legs. Furthermore, by implementing qubit routing in a separate structure from the quantum logic region, qubit sorting can proceed in parallel with the ground state cooling of ions in the logic region, reducing the effective clock-speed of the QPU. Comparisons to the Quantinuum H1~\cite{Pino2020} and Quantinuum H2~\cite{moses2023race} QPUs summarize the cumulative impact of these design choices in the electrical control subsystems (Table \ref{tab:electrode_signals}).

\begin{table}[h]
\centering

\begin{tabular}{lccc}
\hline\hline
System & Num. & Num. & Signals/Qubit \\ 
 & Electrodes & Signals & \\
\hline
H1 & 198 & 198 & 9.9 \\
H2 & 376 & 268 & 4.8 \\
Helios & 1228 &  273 & 2.8 \\
\hline\hline
\end{tabular}

\caption{\label{tab:electrode_signals} The number of electrodes and independent voltage signals per qubit for three different generations of Quantinuum QPUs.}
\end{table}

\subsection{Ion Species - qubit and coolant} \label{sec:ba_qubit} \vspace{0.5em} 

Helios is the first quantum computer to utilize $^{137}$Ba$^+$. We define $\ket{F=1,m_f=0}$ and $\ket{F=2,m_f=0}$ hyperfine levels in the $^{137}$Ba$^+$ electronic ground state as $\ket{0}$ and $\ket{1}$ respectively. The optical transitions used to implement quantum operations are in the visible part of the wavelength spectrum, allowing for laser and optical components that are more mature, reliable, and cost-effective and enables fundamentally better performance. Using more available laser power with better phase performance, we can suppress the leading sources of errors in logic gates, including spontaneous emission errors, laser phase fluctuations, and higher-order Lamb-Dicke errors~\cite{Sorensen2000}. 

Specifically, the single-qubit (1Q) and two-qubit (2Q) gates are implemented with pairs of 515 nm laser beams separated by the $\sim$8.04 GHz qubit frequency splitting. The 1Q gates, $U_{1Q}(\theta,\phi) = e^{(-i \theta/2)(\cos{\phi}X+\sin{\phi}Y)}$, are implemented with co-propagating laser beams for improved phase stability of the Raman interaction and minimal sensitivity to the ions' thermal motion. 1Q Z-rotations, $R_Z(\theta)=e^{-i Z\theta/2}$, are implemented by phase changes in software. The 2Q gates are implemented with beams intersecting the quantum logic zones at 90 degrees to each other such that the difference k-vector is parallel to the crystal axis (Fig.\,\ref{fig:MainFig}e). The 2Q gate protocol is based on the M\o lmer-S\o rensen interaction using wrapper pulses to remove optical phase sensitivity \cite{Pino2020,Lee05}, yielding a native 2Q gate $R_{ZZ}(\theta)=e^{-iZZ\theta/2}$. The gate angle $\theta$ is specified by the user and is varied by adjusting the detuning and duration of the gate. Gate infidelities have been shown to improve for smaller angles \cite{moses2023race}, but here we only benchmark the perfect entangler $R_{ZZ}(\pi/2)$.

\begin{figure}[t]
\centering
{\includegraphics[width=\columnwidth]{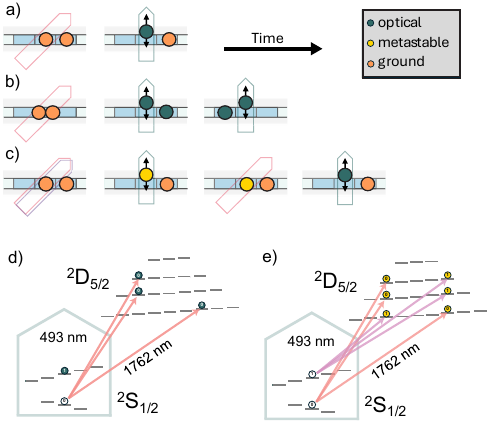}}\\
\caption{Three types of measurements are available in all 8 quantum operation zones which make use of optical ($^2S_{1/2}$, $^2D_{5/2}$), metastable ($^2D_{5/2}$), and ground state ($^2S_{1/2}$) superpositions. All measurements are made with the target ion displaced from the RF null to reduce stray light interacting with non-measured ions~\cite{Gaebler2021crosstalk} as shown with double arrows. (a) Standard measurement occurs when the user specifies a measurement but not for all the qubits in the batch. (b) Protected measurement occurs when the compiler detects an entire batch of qubits will be measured, such as at the end-of-program measurement. Protected measurement performs the $^{2}D_{5/2}$ mapping operations on both qubits prior to state detection such that crosstalk from 493 nm detection light does not affect the measurement outcome. (c) User specified ternary measurement allows the user to obtain a result of 0, 1, or $L$, where $L$ indicates leakage out of the qubit manifold. In this case, each qubit state amplitude is mapped to different parts of the $^{2}D_{5/2}$ manifold~\cite{Senko2025} and any remaining population in the $^2S_{1/2}$ population (representing leakage errors) is measured via induced fluorescence with the 493 nm and 650 nm lasers. Afterwards, a series of pulses independently maps each state amplitude back into the $^2S_{1/2}$ and $^2D_{3/2}$ manifolds allowing measurement of the qubit state (0 or 1). Ternary and protected measure can be combined when an entire batch is measured. (d) Energy level diagram for $^{137}$Ba$^+$ with $^2S_{1/2}$ ground state manifold used for storage and quantum operations, optical superpositions of $^2S_{1/2}$ and $^2D_{5/2}$ are used during standard measurement. An optical superposition shelves the qubit ground state $\vert 0 \rangle$ state to the $^2D_{5/2}$ manifold for measurement. (e) Metastable superpositions in the $^{2}D_{5/2}$ are used for ternary measurements by shelving both ground state qubit levels to detect leakage errors.}
\label{fig:MeasurementSchemes}
\end{figure}

State preparation and measurement (SPAM) are achieved in $^{137}$Ba$^+$ with a combination of lasers at 493 nm, 614 nm, 650 nm and 1762 nm via narrow-band optical pumping Ref.~\cite{an2022high, ransford2021weak}.  The 1762 nm laser is locked to a narrow linewidth cavity to facilitate high-fidelity mapping pulses between the $S_{1/2}$ ground state and $D_{5/2}$ state (Fig.~\ref{fig:MeasurementSchemes}). The standard measurement protocol first maps the $\ket{F=1,m_f=0}$ qubit state to the $D_{5/2}$ manifold with multiple $\pi$ pulses to different levels in $D_{5/2}$. Then the $493$ nm and $650$ nm lasers are turned on to induce fluorescence from all $S_{1/2}$ states. 
Additionally, the 1762 nm laser is used to protect neighboring qubits from measurement crosstalk errors (Fig.~\ref{fig:MeasurementSchemes}b) and enables a ternary (three outcome) measurement to detect leakage population (Fig.~\ref{fig:MeasurementSchemes}c) without the use of ancillas or 2Q gates~\cite{LeakagePatent, sotirova2024high, Allcock2021OMG}.

The QCCD architecture relies on mid-circuit recooling of ions, achieved here with sympathetic cooling applied to $^{171}$Yb$^+$ ions co-trapped with the $^{137}$Ba$^+$ qubit ions. The $^{171}$Yb$^+$ ion is chosen because of similar mass to $^{137}$Ba$^+$ and for the established and straightforward methods for qubit control and state measurement~\cite{PhysRevA.76.052314}. The cooling is performed with lasers tuned near the $S_{1/2}$ to $P_{1/2}$ transition of $^{171}$Yb$^+$ at 369 nm. 
 
To load ions into the QCCD, we photoionize both species from cold atomic beams produced by an atomic source similar to Ref.~\cite{moses2023race}, based on a neutral atom magneto-optical trap (MOT) \cite{De2009BaMOT, johansen2022fast}. Other hardware details, including implementation of all quantum operations are described in the Appendix.

\subsection{QCCD operation} \label{sec:TransportAndParallelOps}\vspace{0.5em}

\begin{figure*}[!t]
    \centering
    \includegraphics[width=1\linewidth]{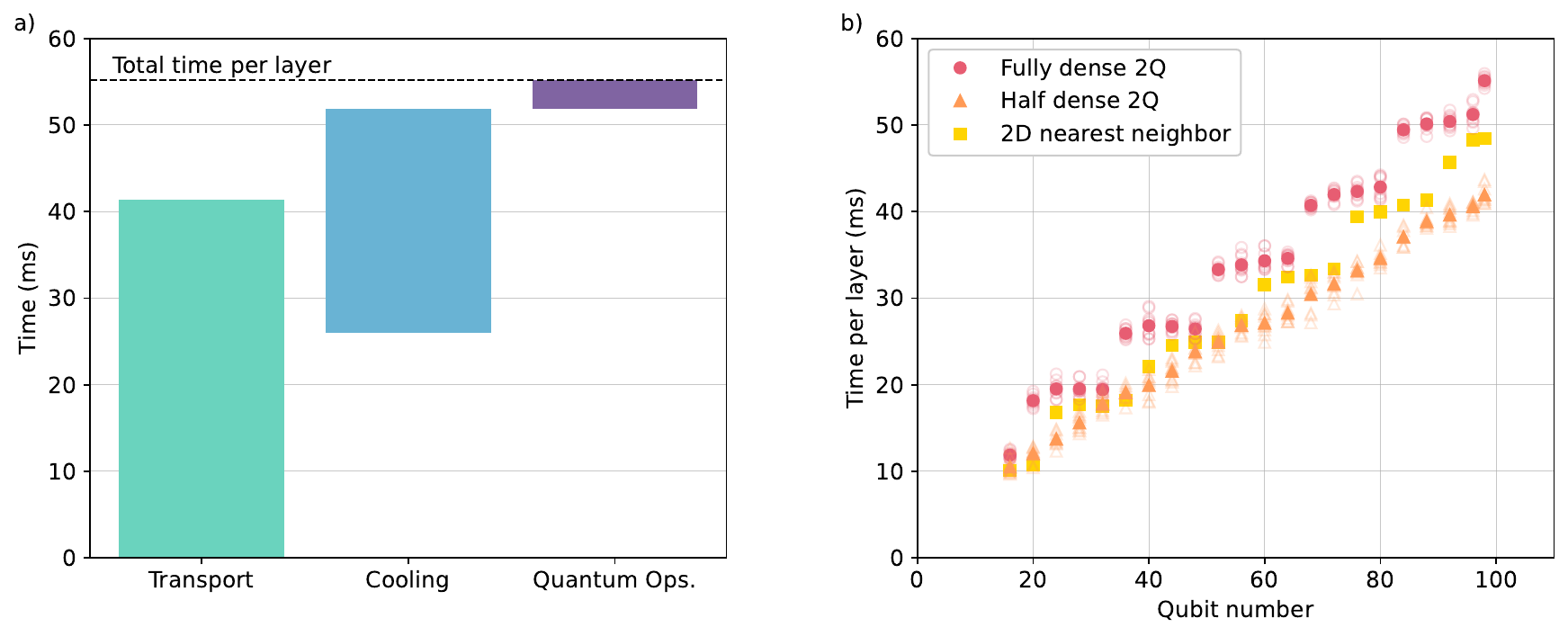}    
    \caption{(a) Time budget per layer for an example depth-10 random program that executes 1Q and 2Q gates on all 98 qubits after an arbitrary permutation each layer, broken down into three categories: ion transport; ground-state cooling; and quantum operations (1Q and 2Q gates). (b) Total time per layer versus number of active qubits for three programs: a random program with fully dense 2Q gates, the same random program with approximately half the 2Q gate density, and a program with 2D nearest-neighbor 2Q gate pairing. For the two random programs, solid points represent the mean of 10 program instances; hollow points show the individual values.}
    \label{fig:depth1_time}
\end{figure*}

In this section, we describe how Helios executes quantum programs using the operations depicted in Fig.~\ref{fig:MainFig}. An arbitrary quantum program is decomposed into ion transport and quantum operations. These operations are not pre-planned but instead executed with a new real-time and dynamic classical control software called ``Helios runtime", which is described in detail in Sec.~\ref{sec:compiler}.

Ions move through the trap using transport operations from four categories: shift, split/combine, junction transport, and rotate. Shift operations translate ions along linear sections in the cache, quantum logic, and leg storage regions. These operations can move both two-ion~Ba$-$Yb~(BY) and four-ion~Ba$-$Yb$-$Yb$-$Ba~(BYYB) crystals. Split (combine) operations separate (merge) BYYB (BY and YB) crystals in the eight operation zones. Junction exit (enter) operations move crystals from (into) the junction into (from) the desired leg in the cache with the desired order, BY or YB. Rotate operations collectively move crystals in the ring clockwise or counterclockwise.

Programs use these transport operations to move qubits between the memory and processor regions of the trap. This cycle occurs during a single layer in a program, in which qubits are removed from ring storage, processed in batches within the quantum logic region, and then returned to ring storage. Every program begins with qubits in a default configuration: 8 BYYB crystals in the quantum logic region and 82 BY crystals in ring storage. Each layer contains up to 7 batches, with a maximum of 16 qubits per batch. 

Using appropriate ion-to-qubit assignments, quantum operations immediately begin on the qubits already in the eight operation zones with individual addressing operations occurring first: state preparation (or reset), 1Q gates, and measure operations. Next, if 2Q gates are required, the BY and YB pairs associated with each zone are combined to BYYB crystals and ground-state cooling begins. In parallel with cooling, qubits for the next batch of gating are moved from the ring storage to the cache. This parallel sorting with ground state cooling allows cooling and gating cycles to run nearly continuously, as the next batch of qubits is ready to shift in as the current batch finishes operations. %Cooling currently lasts approximately 3 ms.

Unlike 1Q, reset, and measure operations, 2Q operations are executed in only four of the eight quantum logic zones (second and fourth zones on top and bottom legs as shown in green in Fig.~\ref{fig:MainFig}b,e). To perform 2Q gates on all 8 qubit pairs, the qubits are first merged and cooled as 8 four-ion crystals in all operation zones and then 2Q beams are applied in the four 2Q operation zones. Immediately after executing the 2Q gates, the four-ion shift operation moves all crystals over by one zone (the crystals in the right edge operational zones are split to BY and YB pairs and then shifted into the storage legs). We then apply a small ($\sim$300 $\mu$s) additional amount of cooling to remove any energy gained from the shift operation and then gate the remaining four crystals. The 2Q gate operation itself requires approximately $\sim$70~$\mu$s to execute.

After executing quantum operations, a batch is complete: its qubits move to leg storage, while qubits in the cache shift to the quantum logic region. This process repeats until all qubits requiring operations have been processed. Lastly, all qubits move from leg storage to the ring, and the cycle begins for the next layer.

Fig.~\ref{fig:depth1_time}a shows timing estimates and a breakdown of operations per layer for a representative program on Helios. The program is constructed as a sequence of 10 layers, in which qubits are randomly paired and receive 1Q and 2Q gates each layer. We define the ``depth-1 time" as the time required to perform the random pairing and 1Q and 2Q gates in a single layer, and use this time as our characteristic figure of merit for processor speed. We estimate the average depth-1 time by measuring the duration of the depth-10 program and dividing it by the number of layers to average any fortunate sort cases, resulting in an average of 55 ms per layer. To illustrate how program details such as 2Q gate density and qubit connectivity impact depth-1 time, we present timing results in Fig.~\ref{fig:depth1_time}b for three example programs as a function of the number of active qubits (for more details, see App.~\ref{sec:appendix_circuit_profiling}).

\subsection{Real time compilation of sorting and gates} \label{sec:compiler}

To realize the full capability of the Helios QPU, the system must be capable of executing arbitrary quantum programs efficiently, including dynamic quantum programs. Optimal decision making for dynamic quantum programs requires a new classical control hardware unit and software compilation stack. This new stack both allows for real-time qubit routing decisions and  increases the level of abstraction of quantum programs---mirroring the way classical computers advanced from writing assembly code to writing high-level programs. 

In particular, Helios is the first trapped-ion QPU to translate operations on a program's ``virtual qubits''~\cite{QASM3SPECTYPES} into operations on corresponding physical qubits on the device in real time---that is, while the program is executing and quantum state is live. This is enabled by the Helios runtime, whose responsibility is to efficiently map virtual qubits to physical qubits on the device and turn declarative gates on virtual qubits into operations on physical qubits.  This runtime enables state-of-the-art user programming constructs for use on a quantum computer (functions that can allocate and de-allocate qubits depending on the control flow of the program), early termination of programs based on mid-circuit measurement or arbitrary classical logic, and classical control flow such as \lstinline{if-then-else} statements, \lstinline{for} loops, and \lstinline{while} loops. This is in stark contrast to the way most gate-level quantum programs, commonly referred to as ``dynamic circuits" ~\cite{qirpaper}, are written right now---as a flat series of gates with conditional gates conditioned on measurements. Many of the Guppy~\cite{Koch2024guppy} programs for the applications discussed in Sec.~\ref{sec:benchmarking} use some of these features. Additionally, any programming language compiling to QIR~\cite{qir_github} such as Q\#~\cite{Svore_2018}, qiskit~\cite{qiskit2024}, OpenQASM 2.0/3.0~\cite{cross2017openquantumassemblylanguage, Cross_2022}, cirq~\cite{Cirq_Developers_2025}, and CUDA-Q~\cite{cudaq} can use QIR adaptive profile features to implement these control flow constructs for programs executing on Helios.

 An example of high-level operations enabled by the Helios runtime is the ``gate streaming'' used in \cite{cr_tbr}. In the Guppy program executed on Helios for this work, a section of the program performs a remote-procedure-call out to a  classical server that is separate from the control system but which is allowed to communicate to the control system via a networking interface~\cite{gate_streaming}. The information transmitted to the control system by the classical server is the measurement basis for each qubit. If a qubit needs no change in measurement basis then the runtime receives no 1Q gate to apply before measurement. In the case that a whole row of BY or YB crystals on the top or bottom legs needs no basis change, the Helios runtime will not perform any extraneous transport to address these qubits. Importantly, this reduces the overall shot time, improving the critical latency times in that application. Efficient gate streaming would be impossible without the real-time identification of qubits provided by the runtime.

The core responsibilities of the Helios runtime are the following: \begin{enumerate*}[label=(\arabic*)]
    \item receive qubit allocation requests on virtual qubits and resolve them to physical qubits;
    \item receive gating requests on allocated virtual qubits; 
    \item transform requested gates on sets of virtual qubits into parallel operations on as many physical qubits as can fit in the quantum operation zones; and
    \item transport batches of physical qubits from the ring into these zones, referred to as a ``sort''.
\end{enumerate*}

Responsibility (1) is performed using a model of the physical QPU state as the program runs and determining efficient mappings from virtual qubits to physical qubits. Responsibilities (2) and (3)  are performed by identifying which quantum logic operations can be done in parallel by storing them in sets contained in a data-structure we refer to as a ``slice''. Sequences of slices are accumulated into another data-structure that drives the sorting of each slice to execute the quantum logic operations within. Responsibility (4) is performed by doing an $O(n)$ traversal over the ring storage to determine which two pairs in a slice have qubits closest to the cache. The runtime then assigns one pair to move to the top leg and the other to the bottom. Subsequently, the algorithm determines the smallest number of rotations needed to move the two pairs into BYYB crystals in both legs. This process is visualized in Fig.~\ref{fig:MainFig}. This process repeats until either enough pairs are moved into the cache to fill a batch, or until no more pairs need to be sorted. Finally, the runtime dispatches the calculated sort by generating these operations as a queue of commands to lower-level control system software for performing transport operations and parallelized cooling as outlined in~\ref{sec:TransportAndParallelOps}. After all of the quantum logic operations have been executed in a given slice via repetitions of this sort, transport is generated to return the qubits back into the ring storage--and the sorting algorithm repeats for subsequent slices.

\section{Benchmarking} \label{sec:benchmarking}

\subsection{Overview}

To see how Helios performs in practice and understand current limitations,
we characterize individual operations with component-level benchmarks and full-device operation with system-level benchmarks~\cite{moses2023race}.
Operations include SPAM, 1Q and 2Q gates,
mid-circuit measurements and resets (MCMRs),
and qubit idle during ion transport. We perform two separate system-level benchmarking experiments~\cite{Cross2019,BlumeKohout2020, Wack2021, Tomesh2022, Lubinskit2023,Proctor2024}, both of which are examples of volumetric benchmarks~\cite{BlumeKohout2020}. The first involves random Clifford circuits with MCMR, which can be simulated classically.
We include MCMRs,
unlike most prior work,
because they are necessary for quantum error correction.
The second experiment is mirror benchmarking of random circuit sampling (RCS), which is an appealing benchmark because the
quantum computational power can be measured by the classical simulation cost. The use of mirroring allows for estimating the circuit fidelity where classical simulation is unfeasible.

In the following, we first describe the component-level benchmarks in Sec.~\ref{sec: component benchmarks}, with a summary of the results given in Tab.~\ref{tab:component benchmarks}.
We then present our system-level benchmark results with a detailed comparison to the prediction from the component-level benchmarks in Sec.~\ref{sec: system benchmarks}.

\begin{table}[h]

\centering
\resizebox{\columnwidth}{!}
{
\begin{tabular}{llr}
\hline\hline
Component & Metric & Value ($\times10^{-4}$) \\
\hline
SPAM (standard) & Average error & 4.8(6)\\
SPAM (ternary) & Average error & 17(1)\\
1Q gates & Clifford avg. infidelity & 0.25(1) \\
2Q gates & Avg. infidelity (2QRB) & 7.9(2) \\
2Q gates & Avg. infidelity (CB) & 8.1(2)\\
Transport idle & Linear memory error rate & 5(1)\\
Transport idle & Quadratic memory error & 0.7(2)\\
&parameter&\\ 
MCMR crosstalk & Avg. infidelity (global) & 0.48(1)\\

\hline\hline
\end{tabular}
}
\caption{\label{tab:component benchmarks} Component-level benchmark values,
averaged over all operation zones.}
\end{table}

\subsection{Component-level benchmarks}\label{sec: component benchmarks}

\subsubsection{State-preparation and measurement}
It is difficult to differentiate state preparation errors from measurement errors \cite{christensen2020high}, although from detailed modeling of $^{137}$Ba$^+$ qubits we expect state preparation errors to be the largest contributor~\cite{an2022high}.

We measure SPAM errors by preparing 16 qubits in the 8 operation zones in the $\ket{0}$ or $\ket{1}$ states,
and measuring each qubit.
For any given shot, the state preparations are randomized among the different qubits,
but we ensure that each qubit is prepared in each state for the same total number of shots.
We run two experiments: standard measurement that ideally differentiates $\ket{0}$ from $\ket{1}$ but falsely returns $\ket{1}$ in the event that the qubit has leaked,
and a ternary measurement, shown in Fig.~\ref{fig:MeasurementSchemes}c, that ideally differentiates $\ket{0}$, $\ket{1}$, and leaked states.
For both experiments, we take 4000 shots per state preparation.

For the standard measurement, we measure errors of $8(1)\times10^{-4}$ and $1.6(5)\times10^{-4}$ when preparing $\ket{0}$ and $\ket{1}$, respectively.  
Because this measurement protocol mistakenly detects leaked states as $\ket{1}$, the reported error for preparing and measuring $\ket{1}$ will not catch all errors \cite{an2022high}.
For the ternary measurement, we find an average leakage probability of $4.2(7)\times10^{-3}$,
and in the event of non-leakage we measure SPAM errors of 
$7(1)\times10^{-4}$ and $2.8(2)\times10^{-3}$,
for $\ket{0}$ and $\ket{1}$, respectively.
Although the ternary measurement reveals more information as it can detect leakage,
it also has a larger SPAM error due to a larger number of shelving pulses involved.
The SPAM errors reported in Tab.~\ref{tab:component benchmarks} are averaged between the two state preparations.

\begin{figure*} % 
    \includegraphics[width=\textwidth]{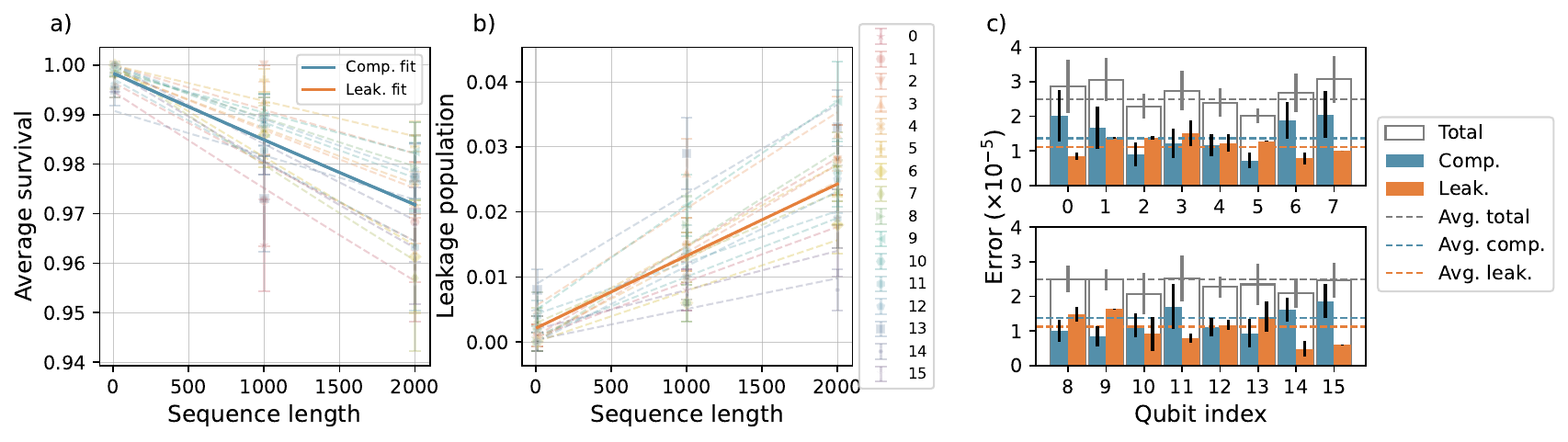}    
    \caption{1QRB data:
    (a) 1QRB survival probability as a function of sequence length, for the 16 qubits occupying the 8 operation zones.
    (b) 1QRB measured leakage population as a function of sequence length.
    The leakage rate is combined with the survival decay rate to compute the 1Q Clifford infidelity.
    (c) Breakdown of 1Q Clifford error rates into computational (comp.) errors and leakage (leak.) errors,
    for the 16 individual qubits. Label locations correspond to qubit locations in Fig.~\ref{fig:MainFig}b with qubits 0-7 in the top operation zones and 8-15 in the bottom (two per zone ordered left to right).}
    \label{fig:SQRB}
\end{figure*}

\subsubsection{Single-qubit gates}\label{sec: SQRB}
Single-qubit gate errors are primarily caused by spontaneous emission during the Raman gate, laser phase and intensity noise, and finite qubit coherence. Importantly, spontaneous emission causes leakage outside of the computational subspace. We quantify 1Q gate errors by Clifford randomized benchmarking (RB)~\cite{Magesan12}, with details provided in App.~\ref{sec: benchmarking data appendix}.

We follow the methods in Ref.~\cite{chen2025} to account for leakage in the 1Q infidelity estimate. The ternary measurement allows us to measure the leakage population at the end of every circuit without the use of ancilla qubits (as was done in Ref.~\cite{moses2023race}).
We estimate the rate of leakage per 1Q Clifford $r_{L}$
by the rate at which the measured leakage population increases with sequence length.
The probability of observing the expected computational state decays exponentially
due to non-leakage errors as $p(l)=A(1-r)^l+1/2$ for sequence length $l$.
The reported 1Q error is the Clifford average infidelity $\epsilon_{avg, 1Q}=r/2+r_L$~\cite{chen2025}.

Figure~\ref{fig:SQRB} shows the survival probability and the leaked population as a function of $l$,
for all 16 qubits in the 8 operation zones.
We obtain a zone-averaged 1Q error of $2.5(1)\times10^{-5}$,
which includes a leakage rate of $1.12(6)\times10^{-5}$.
The error bars represent a 1-sigma confidence interval obtained from bootstrapping~\cite{Efron1993}.
The leakage rates and infidelities for each individual qubit are given in Tab.~\ref{tab: SQRB data}. The measured errors can be compared with our predictions from physical error models of $2.6(6)\times10^{-5}$ that account for measured laser intensity noise, calculated spontaneous emission, and measured memory error.

%In principle, errors in the individual 1QRB experiments may not be independent.
Finally, we ran a statistical hypothesis test for correlated errors in the simultaneous 1QRB data.
An error channel on multiple subsystems is correlated if it cannot be factored into a tensor product of individual error channels on each subsystem,
and such correlated errors are a signature of crosstalk.
We found no evidence of correlated errors at the 95\% confidence level (see App.~\ref{app:correlation_analysis} for analysis details).

\subsubsection{Two-qubit gates}
\begin{figure*} % 
    \includegraphics[width=\textwidth]{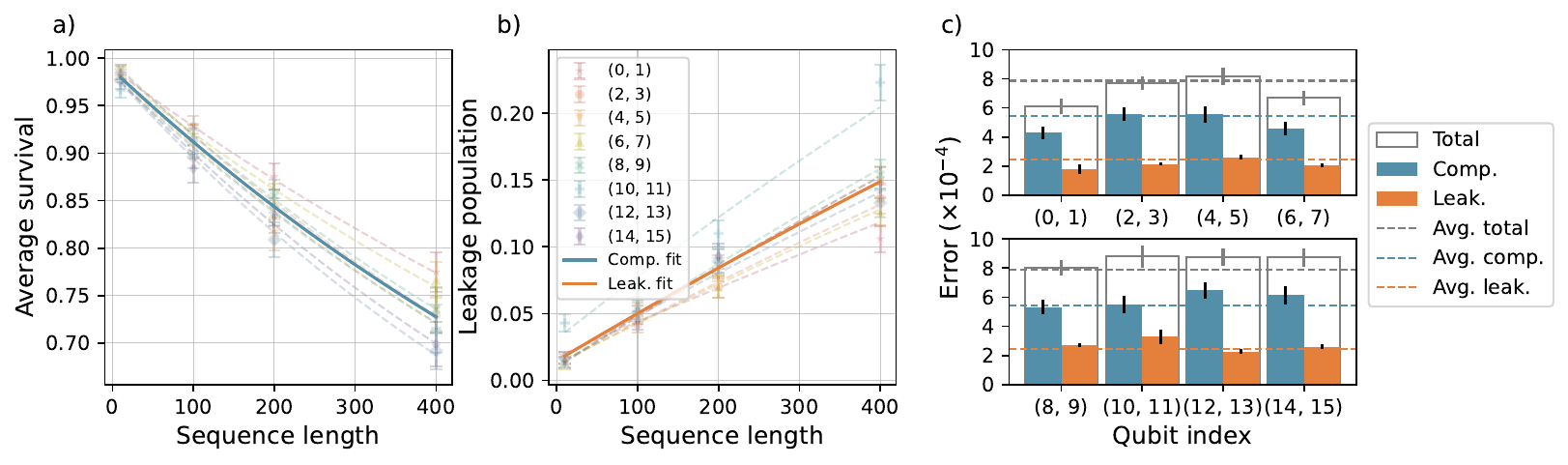}
    \caption{2QRB data: 
    (a) 2QRB survival probability as a function of sequence length for 8 qubit pairs in the 8 operation zones.
    Sequence length here refers to the number of Clifford group elements.
    (b) 2QRB measured leakage rate as a function of sequence length. The leakage rate is combined with the survival decay rate to compute the 2Q infidelity.
    (c) Breakdown of $R_{ZZ}(\pi/2)$ errors into computational and
    leakage errors, for the 8 qubit pairs. Label locations correspond to qubit locations in Fig.~\ref{fig:MainFig}b with qubits 0-7 in the top operation zones and 8-15 in the bottom (two per zone order left to right)
    }
    \label{fig: TQRB decays}
\end{figure*}

\begin{figure*}[t]
\includegraphics[width=\textwidth]{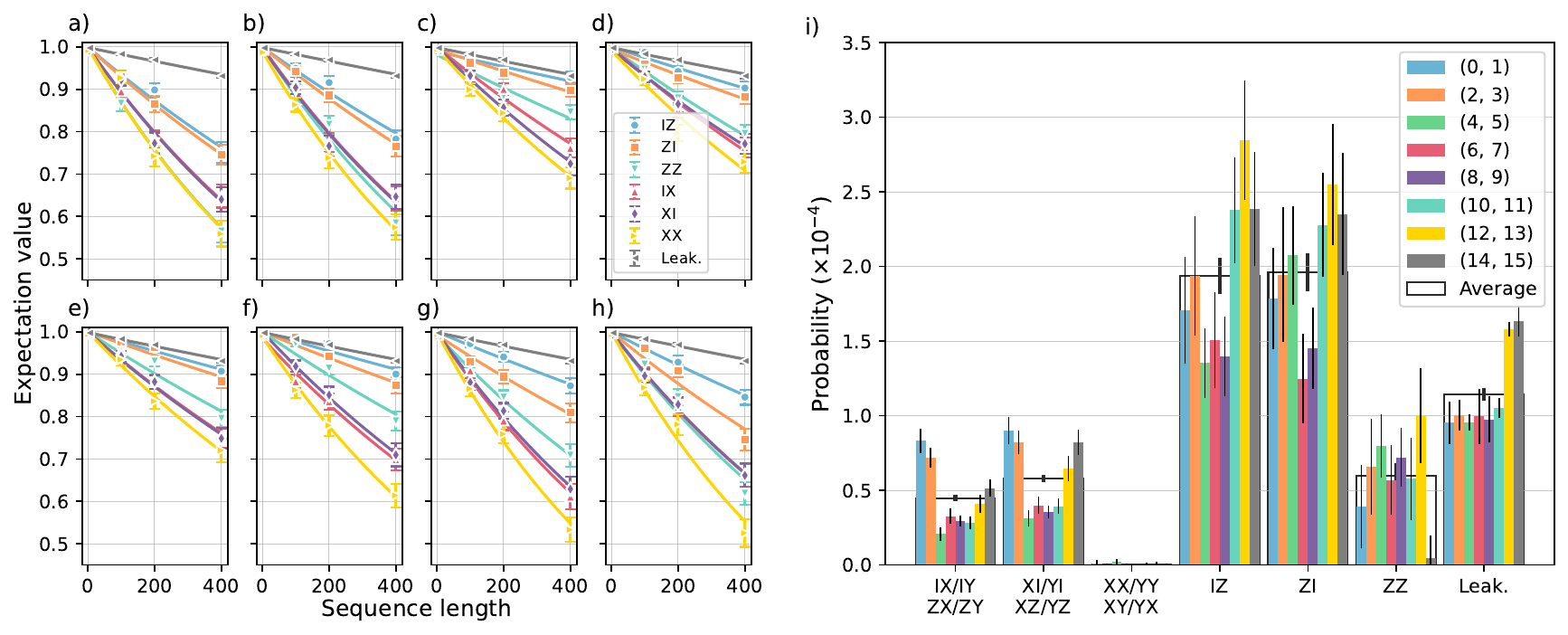}
\caption{2QCB data:
(a-h) Pauli expectation values and measured leakage rate as function of sequence length,
for the 8 operation zones in order of Fig.~\ref{fig:MainFig}
(i) Pauli error probabilities and leakage survival probability for the native $R_{ZZ}(\pi/2)$ gate, up to unlearnable degrees of freedom, for the 8 operation zones.}
\label{fig: CB}
\end{figure*}

Errors in the $R_{ZZ}(\theta)$ gates are caused by spontaneous emission from the Raman lasers and experimental imperfections including laser phase and intensity noise at the ion's position, thermal motion of the ions, voltage noise on the electrodes, and imprecise calibrations of the gate parameters.
We validate the performance of the maximally entangling $R_{ZZ}(\pi/2)$ gate (referred to as the 2Q gate)
using both Clifford 2QRB and cycle benchmarking (CB). 
% We account for leakage using techniques similar to those used in benchmarking the 1Q gates. 
Additional details of each implementation is in App.~\ref{sec: benchmarking data appendix}.

We again follow the methods in Ref.~\cite{chen2025} to account for leakage in the 2QRB infidelity estimate.
The leaked population versus sequence length is
used to extract a leakage rate per Clifford,
which is rescaled into a leakage rate per 2Q gate $r_{L,2Q}$,
using the fact that there are 1.5 2Q gates per 2Q Clifford on average.
We fit the survival probability of the remaining population to the decay model $p(l)=A(1-r)^l+1/4$,
and the average infidelity of the non-leakage error component per Clifford is given by $3r/4$,
which is rescaled into an average infidelity per 2Q gate of $r/2$.
The average infidelity per 2Q gate (including leakage) is then computed as $\epsilon_{avg, 2Q}=r/2+r_{L,2Q}$.
We note that our rescaling of the error per Clifford into an error per 2Q neglects the errors from 1Q gates
and memory errors during the 2QRB sequence,
which we estimate to contribute $1.2(2)\times10^{-4}$ per 2Q gate.

The experimental 2QRB data is shown in Fig.~\ref{fig: TQRB decays}.
We obtain a zone-averaged 2Q infidelity of $\epsilon_{avg, 2Q}=7.9(2)\times10^{-4}$,
which includes a leakage rate of $r_{L,2Q}=2.4(1)\times10^{-4}$.
The leakage rates and infidelities for each individual qubit pair are given in Tab.~\ref{tab: TQRB data}.
The leakage errors arise both from spontaneous emission error,
which we measure to be $1.0(2) \times 10^{-4}$ in agreement with the model of \cite{Moore2023},
and from the leakage memory error (discussed in Sec.~\ref{sec:memory_error}).
In total, we expect leakage to contribute $1.7(2) \times 10^{-4}$ of the error.

Our measured value of $7.9(2)\times10^{-4}$ can be compared to a total expected error per 2Q gate of $3.5(4)\times10^{-4}$,
which we predict from an error budget consisting of
spontaneous emission errors, memory error, and 1Q pulse errors plus other characterized experimental sources of noise such as laser phase and intensity noise,
thermal motion of the ions, and imprecise calibrations.
The discrepancy of the measured 2Q error with predicted value could be explained by a number of factors including higher leakage error in the operational zones due to finite extinction of the resonant detection beams present, non-thermal motional distributions, crosstalk, or other unaccounted for effects.

Just as with the 1QRB data,
we performed a statistical test for the presence of correlated errors in the 2QRB data and found no significant evidence of correlated errors across the qubit pairs (see App.~\ref{app:correlation_analysis} for details).

We also perform two-qubit cycle benchmarking (2QCB)~\cite{Erhard2019} to estimate a partial Pauli error model for the 2Q gate in each operation zone, with the experimental and theoretical details supplied in App.~\ref{sec: benchmarking data appendix}. Fig.~\ref{fig: CB} shows the expectation value decays
and estimated Pauli error channels,
for each qubit pair.
We find the zone-averaged infidelity is $8.1(2)\times10^{-4}$, which includes a leakage rate of $1.14(4)\times 10^{-4}$, and is dominated by $IZ$ and $ZI$ errors.
We note that our estimate of leakage rate per 2Q gate from 2QCB is about a factor of two smaller than the estimate from 2QRB.

\begin{figure*}[t]
\includegraphics[width=\textwidth]{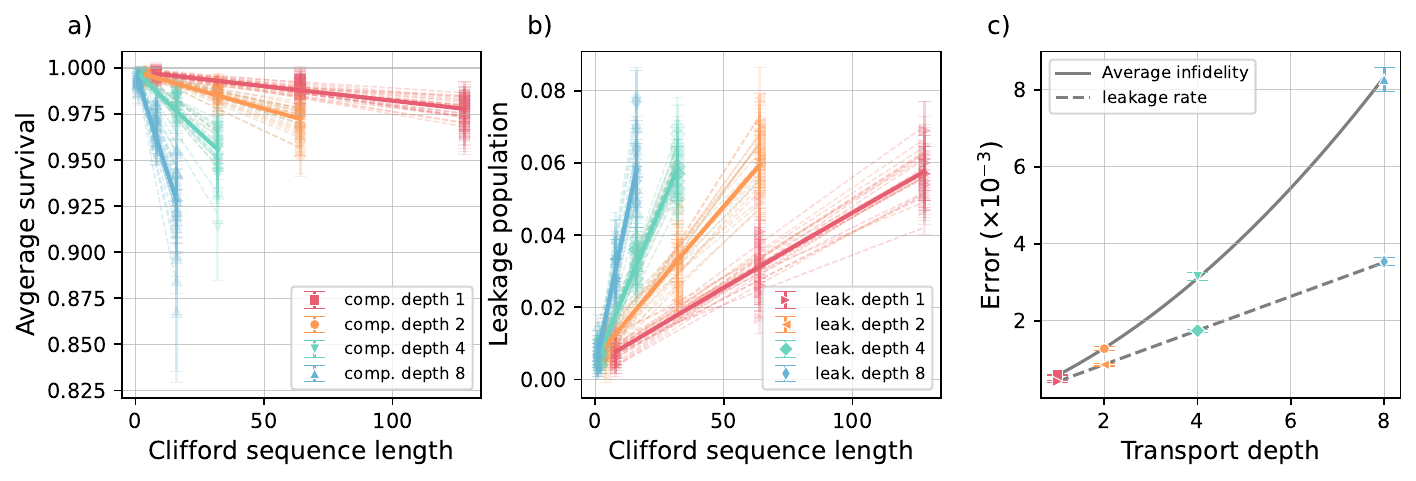}
\caption{Transport-1QRB data:
(a) Survival probability as a function of Clifford sequence length for 98 qubits grouped into 4 groups.
(b) Measured leakage population as a function of Clifford sequence length.
(c) Qubit-averaged leakage rate (dashed curve) and total memory error (solid curve) as a function of the number of depth-1 transport operations.
}
\label{fig: transport SQRB}
\end{figure*}

\begin{figure}[t]
\includegraphics[width=\columnwidth]{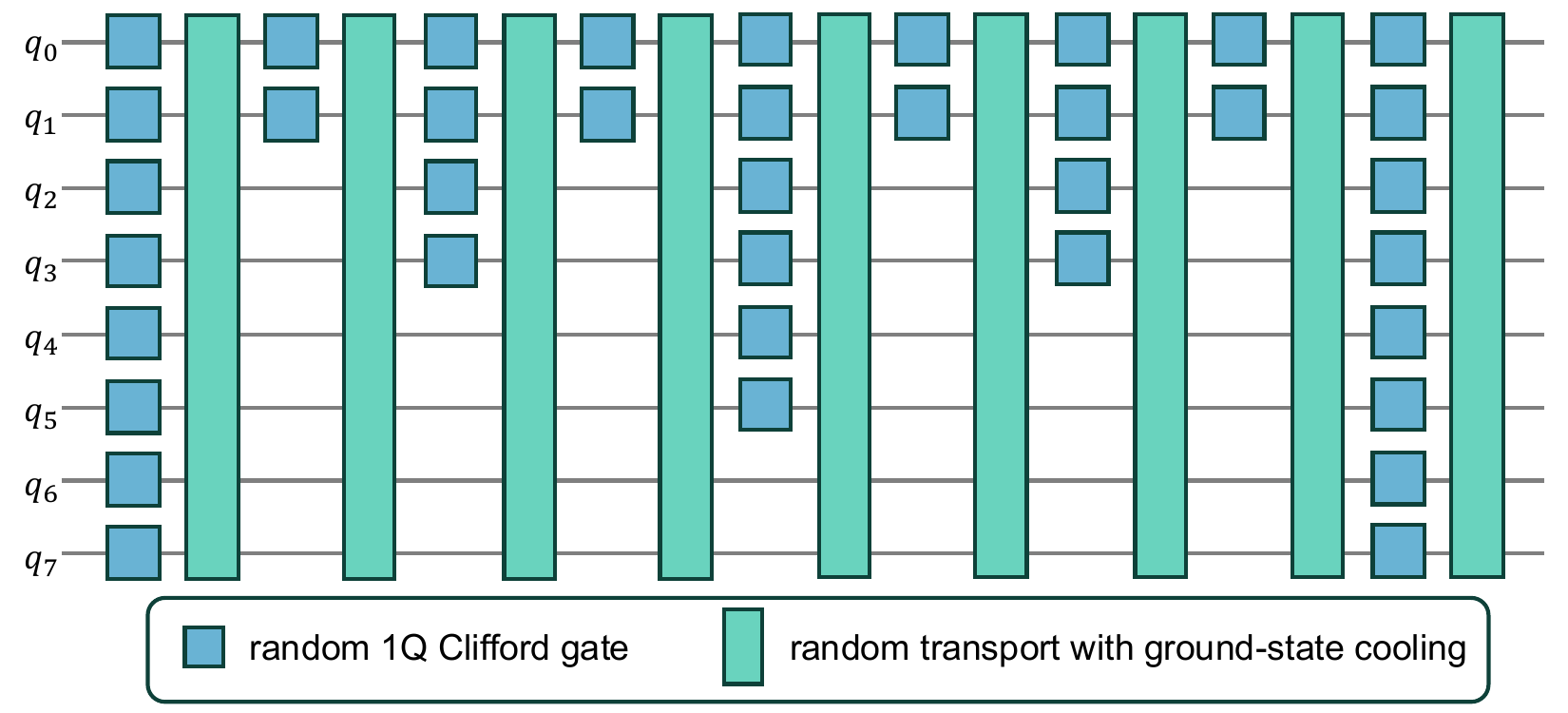}
\caption{Example circuit diagram for transport-1QRB where different qubits receive different number of transport rounds between gates: $q_0$ and $q_1$ have depth-1, $q_2$ and $q_3$ have depth-2, $q_4$, $q_5$ have depth-4, and $q_6$ and $q_7$ have depth-8.}
\label{fig: transport SQRB circuit diagram}
\end{figure}

\subsubsection{Transport idle memory errors}\label{sec:memory_error}

Qubits idle during ion transport and cooling and incur memory errors due to spatiotemporal magnetic field inhomogeneities, with their impact being heavily dependent on the circuit structure and its specific transport schedule.
As a figure of merit we define the depth-$n$ memory error to be the average infidelity per qubit
after randomly pairing all qubits, performing the transport and cooling operations that would be required to apply 2Q gates on all pairs (but no actual gate operations),
and repeating this process $n$ times. 

We measure memory error with a variant of 1QRB that interleaves random transport between 1Q Clifford gates, referred to as transport-1QRB~\cite{Sheldon16,moses2023race}. Our method here differs from Ref.~\cite{moses2023race} in that we partition the 98 qubits into groups
where the qubits in each group have a random 1Q
Clifford operation applied after every $k$ rounds of depth-1 transport operations as shown in Fig.~\ref{fig: transport SQRB circuit diagram}.
The qubits in the different groups will have a different amount of transport and idle time between Clifford operations,
which allows us to extract how memory errors scale with the number of depth-1 transport operations for random circuits.

We run transport-1QRB circuits on the 98 qubits
with one Clifford between every $k\in\{1,2,4,8\}$ transport operations 
Additionally, we use the ternary measurement to extract any leakage errors during transport.
Fig.~\ref{fig: transport SQRB}a and b show the measured decay in transport-1QRB for computational and ternary measurements respectively.
The decay curves are clustered into 4 groups determined by $k$. By fitting the decay curves and accounting for the leakage rate using the same procedure as in Sec.~\ref{sec: SQRB},
we obtain the Clifford infidelity for each qubit.

Fig.~\ref{fig: transport SQRB}c shows a plot of the Clifford infidelity as a function of the number of depth-1 transport operations,
averaged over all qubits in the corresponding group. The expected scaling of memory error with delay time varies depending on the time scale of the noise sources \cite{Sepiol2019}.
For this reason we fit the memory error versus $l$ to a quadratic equation $a+bl+cl^2$ where $b$ and $c$ capture the linear memory error rate (from fast noise) and quadratic memory error parameter (from slow noise), respectively~\cite{Sheldon16}.

\begin{figure*}[t]
\includegraphics[width=\textwidth]{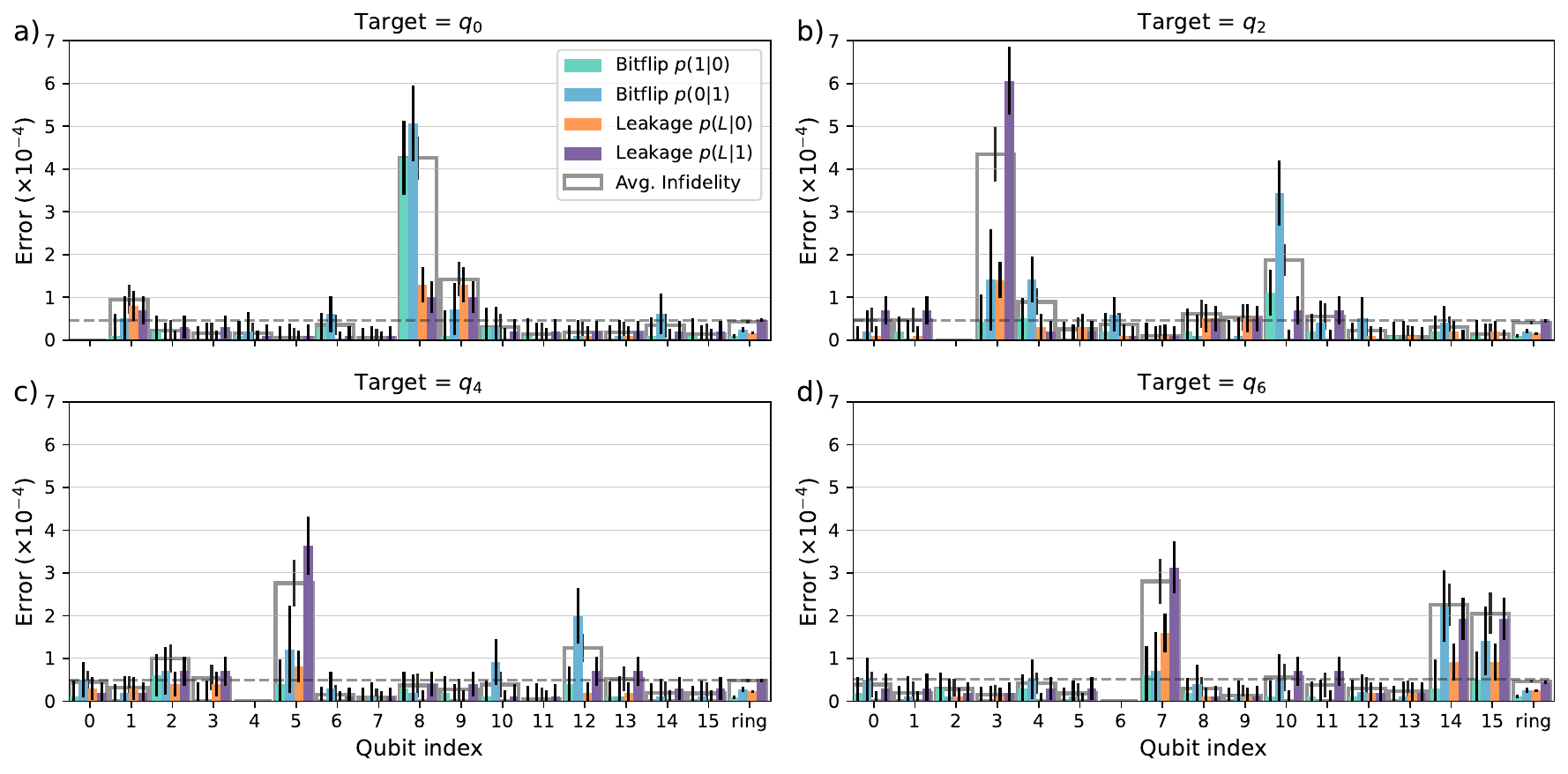}
\caption{MCMR crosstalk data with estimated rates of different error channels (smaller bars) and average infidelity (wider bars) with different individual target qubits (repeatedly MCMR'ed) and spectators (witnesses to crosstalk). We report conditional probabilities $p(i|j)$ for transitioning to state $i$ given state $j$ where $i,j=\{0, 1, L\}$ and $L$ represents the sum of all leaked state populations. The x-axis labels the qubit locations in the quantum operation zones with 0-7 in the top and 8-15 in the bottom (left to right). The ``ring'' bin contains an average over the remaining 82 qubits that sat in the ring during the test.  The dashed horizontal lines indicate the global average infidelity which is the mean crosstalk error for all 97 spectator qubits.}
\label{fig: mcmr crosstalk}
\end{figure*}

From the fit to the data, we infer a linear memory error rate of 
$5(1)\times 10^{-4}$
and a quadratic memory error parameter of $7(2)\times 10^{-5}$.
We find that the leakage error scales linearly with the number of transport operations, with a rate of $4.0(2)\times 10^{-4}$ and accounts for nearly all of the linear memory error. The expected coherent error from typical drift in magnetic fields between calibrations (every $\sim$ 5~s) of approximately $10~\mu$G is $3\times10^{-5}$ in a depth-1 circuit. The remaining coherent error may be explained by imperfections in the phase tracking routine or other unaccounted sources of noise.

\subsubsection{Mid-circuit measurement and reset crosstalk}\label{sec:mcmr}
MCMR causes crosstalk errors on un-measured or un-reset qubits that absorb stray measurement or reset light. The resulting spontaneous emission can lead to bit-flip, leakage, or dephasing errors.  

We measure MCMR crosstalk errors by partitioning the 98 qubits into target qubits that are measured and reset repeatedly. Spectator qubits are prepared in the $\ket{0}$ or $\ket{1}$ and we use the ternary measurement at the end. The combination allows us to differentiate bit-flip rates from leakage rates to get a more detailed picture of the crosstalk error channel. The test was repeated for individual target qubits in the operation zones to illustrate the structure of MCMR crosstalk errors as shown in Fig.~\ref{fig: mcmr crosstalk}. Further details are provided in App.~\ref{sec: benchmarking data appendix}.

It is clear that ions sitting adjacent to the 493 nm lasers applied to the target ion (in the same zone or neighboring zone above/below as shown in Fig.~\ref{fig:MainFig}e) have much larger crosstalk errors.  We distinguish between local (three ions that are laser-adjacent) and global (all 97 spectators) crosstalk, reporting per MCMR average crosstalk infidelities $2.1(1)\times 10^{-4}$ and $4.8(1)\times 10^{-5}$, respectively.  The linear memory error rate (see Table~\ref{tab:component benchmarks}) contributes background leakage at a per MCMR rate of roughly $9(2)\times10^{-6}$ to the measured average infidelities.

\subsection{System-level benchmarks}\label{sec: system benchmarks}

\subsubsection{Random Clifford circuits with mid-circuit measurements}\label{sec: binary RB}

\begin{figure}[t]
\includegraphics[width=\columnwidth]{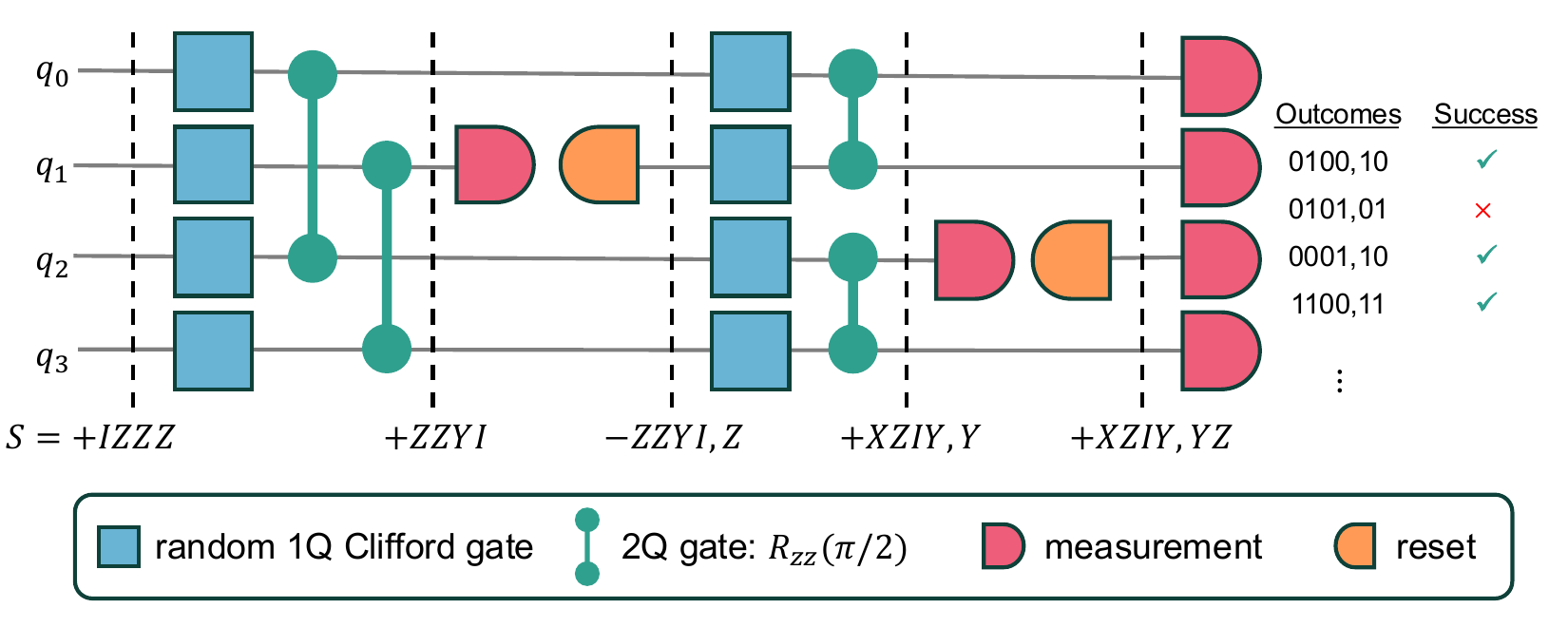}
\caption{Example of a random Clifford circuit with MCMRs,
with parameters $N=4$, $n_m=1$, and $l=2$.
In each layer the qubits are randomly paired and gated and then a random subset of qubits is measured and reset.
An initial random stabilizer $S$ is chosen and evolved through the circuit to determine a binary outcome (success or failure) for each shot.}
\label{fig: binary RB circuit diagram}
\end{figure}

To test the ability of Helios to execute arbitrary 98-qubit circuits using all primitive components,
we run circuits with layers consisting of random Clifford 1Q and 2Q gates and MCMRs.
Ref.~\cite{Hines2024} introduced circuits with random Clifford layers as a scalable system-level benchmark called binary randomized benchmarking (BiRB).
An extension allowing for MCMRs was given in~\cite{Hothem2024}, called quantum instrument randomized benchmarking (QIRB).
Our circuits are constructed similarly to Ref.~\cite{Hothem2024} with a few small modifications.
An example circuit diagram is shown in Fig.~\ref{fig: binary RB circuit diagram}.

In our implementation, a length $l$ circuit on $N$ qubits with $n_m$
MCMRs per layer consists of the following for each layer:
%contains $l$ layers each consisting of the following:
\begin{itemize}
\item A distinct uniformly random 1Q Clifford is applied to each qubit.
\item The $N$ qubits are uniformly randomly paired into $\left\lfloor \frac{N}{2} \right\rfloor$ qubit pairs,
and the 2Q gate $R_{ZZ}(\pi/2)$ is applied to each pair, with Pauli-twirling applied to the 2Q gates.
\item A uniformly random subset of $n_m$ qubits are sampled,
and for each qubit a 1Q Clifford is applied to prepare a measurement in a particular Pauli basis, followed by an MCMR operation. 
\end{itemize}

\begin{figure}[t] % 
    \centering
    \includegraphics[width=0.5\textwidth]{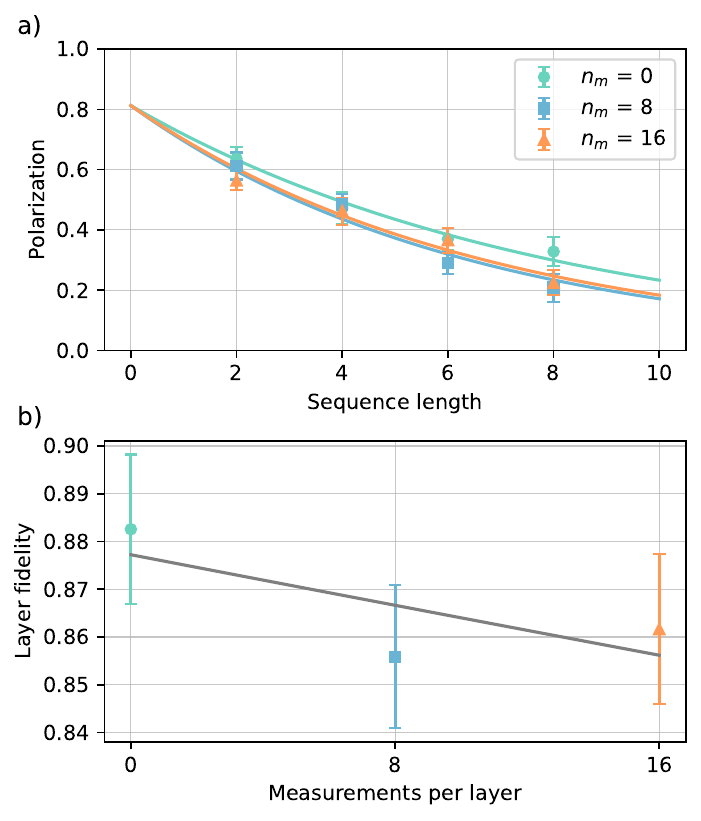}
    \caption{Random Clifford circuits with MCMR data:
    (a) Polarization versus sequence length for different numbers of MCMRs per layer.
    (b) Layer fidelity versus number of MCMRs per layer.}
    \label{fig:binary RB fig}
\end{figure}

To classically verify correct circuit outputs,
we track a random initial stabilizer through the circuit,
as explained in App.~\ref{sec: app: QiRB}.
The parity of the evolved stabilizer defines a success/failure trial.
For the purpose of fidelity estimation, the average success probability is rescaled into a quantity called the polarization~\cite{Hines2024}, defined as $y_{pol}=2p_{succ}-1$.
A polarization of 1 corresponds to perfect success,
whereas a polarization of 0 corresponds to 50\% success, or random guessing.
A plot of $y_{pol}(l, n_m)$ versus $l$ for different values of $n_m$ is shown in Fig.~\ref{fig:binary RB fig}a.
Let $F(n_m)$ be the process fidelity per circuit layer
as a function of $n_m$.
We estimate $F(n_m)$ by fitting the polarization to an 
exponential decay model.

Figure~\ref{fig:binary RB fig}b shows a plot of $F(n_m)$ versus $n_m$.
We note that the layer fidelity actually increases slightly (with overlapping error bars) as $n_m$ increases from 8 to 16.
This is explained by the fact that a batch of 16 measurements in the operation zones utilizes the protected measure scheme (explained in Fig.~\ref{fig:MeasurementSchemes}b),
which protects against MCMR crosstalk in the operation zones.

To see whether the results are consistent with our component benchmarks,
we first compute an effective 2Q gate error $\epsilon_{\mathrm{eff},2Q}$
from the $n_m=0$ data, using
\begin{equation}
F(n_m=0) = (1-5\epsilon_{\mathrm{eff},2Q}/4)^{\left\lfloor \frac{N}{2} \right\rfloor},
\end{equation}
where the factor of $5/4$ comes from the conversion between process and average fidelity~\cite{Nielsen2002}.
The effective 2Q gate error includes errors from 2Q gates,
1Q gates, and memory errors, and it can be thought of as the infidelity of a 2Q depolarizing channel that would best fit the data in the absence of all other errors.
We find $\epsilon_{\mathrm{eff},2Q}=2.0(3)\times10^{-3}$,
whereas an accounting of 2Q and memory errors from Tab.~\ref{tab:component benchmarks} predicts $2.2(1)\times10^{-3}$ (see Sec.~\ref{sec: app: QiRB} for details).

We next compute effective MCMR errors $\epsilon_{M}$ for the $n_m=8$ and $n_m=16$ data,
using the heuristic formula
\begin{equation}
F(n_m) = (1-5\epsilon_{\mathrm{eff},2Q}/4)^{\left\lfloor \frac{N}{2} \right\rfloor}(1-3\epsilon_{M}/2)^{n_m}
\end{equation}
together with our computed value of $\epsilon_{\mathrm{eff},2Q}$.
We find $\epsilon_M(n_m=8)=(2.6\pm1.3)\times10^{-3}$
and $\epsilon_M(n_m=16)=1.0(7)\times10^{-3}$.
By comparison, adding the component-level SPAM error and the MCMR crosstalk error,
we predict effective MCMR errors of
2.2(1)$\times10^{-3}$ and $1.7(1)\times10^{-3}$ for $n_m=8$ and $n_m=16$
(see Sec.~\ref{sec: app: QiRB} for details).
We conclude that the data from our random Clifford with MCMR circuits is consistent with our measured component-level 2Q error,
but tighter error bars are needed to assess
the consistency of the effective MCMR errors.
We remark that our method of comparison is heuristic and a rigorous methodology for comparing component-level to system-level benchmarking performance is an open problem.

\subsubsection{RCS mirror benchmarking}

Random circuit sampling (RCS) is a system-level benchmark assessing how effectively a quantum computer can generate computationally complex quantum states \cite{arute2019}. Like BiRB, RCS probes the extent to which quantum circuits obtain the performance expected from component-level benchmarks. At the same time, because the classical difficulty of sampling from the outputs of random quantum circuits has been extremely well-studied over the last decade \cite{RevModPhys.95.035001}, RCS provides a well-vetted benchmark for the computational power of a quantum computer.

\begin{figure}[!t] % 
    \centering
{\includegraphics[width=\columnwidth]{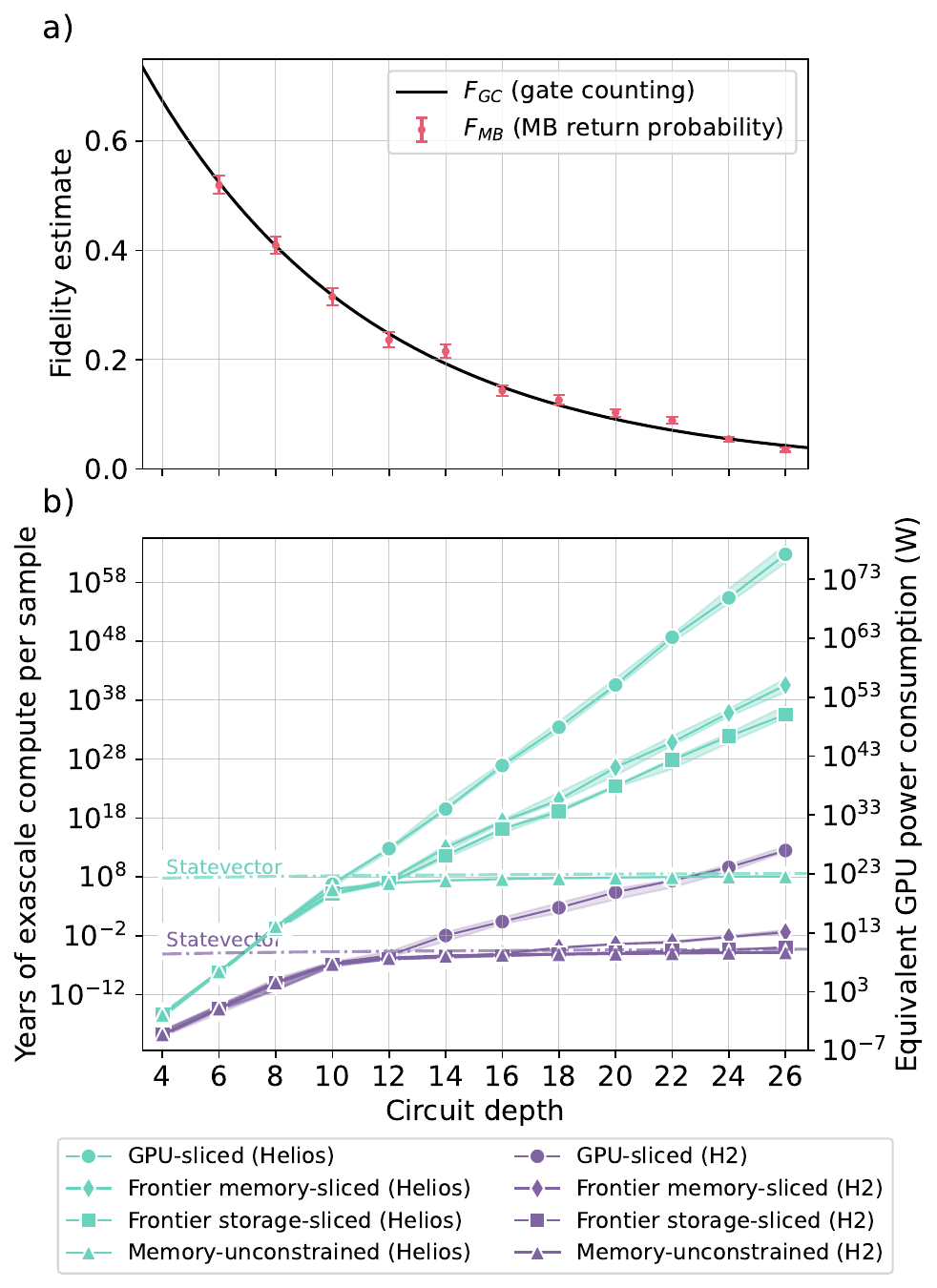}}
    \caption{(a) Fidelity of $N=98$ mirrored RCS circuits as a function of circuit depth (red). The best-fit gate-counting curve is overlaid (black), demonstrating consistency with an exponential decay with depth. (b) Estimated cost of classical sampling via tensor-network contraction from RCS circuits of varying depth on both H2 (purple) and Helios (teal). The left axis reports time in years required to draw a single sample by tensor-network contraction on a state-of-the-art supercomputer (achieving about $10^{18}$ FLOPs per second). The right axis shows the required power (assuming state-of-the-art GPU power efficiency of roughly $10^{11}$ FLOPs/W) to perform contraction-based sampling at the same rate that Helios can draw samples. Costs are quoted across different assumptions on the total memory footprint of the contraction (in a similar fashion to \cite{Abanin2025}), corresponding to the \texttt{cotengra} contraction width $\mathcal{W}$. Triangles show costs assuming access to unlimited memory ($\mathcal{W} = \infty$), which saturates at large depths to the $\sim 2^N$ scaling of statevector simulation; squares ($\mathcal{W} = 54$) allow use of all external storage of the Frontier supercomputer, while diamonds ($\mathcal{W} = 49$) restrict to the available memory on Frontier, and circles ($\mathcal{W} = 30$) correspond to spreading the slices independently among state-of-the-art GPUs. Shaded bands indicate the range of costs obtained over 5 random circuit instances at each depth, with the markers indicating the median cost.}
    \label{fig:rcs}
\end{figure}

Leveraging the arbitrary connectivity of the Helios quantum computer, we consider RCS with circuit geometries constructed from colorings of random-regular graphs \cite{DeCross2025}: A layer depth-$l$ random circuit is constructed by interleaving $l$ layers of 2Q $R_{ZZ}(\pi/2)$ gates (each layer containing $N/2$ 2Q gates) with $l+1$ layers of Haar-random 1Q gates (each layer containing $N$ 1Q gates). While the fidelity of such circuits can in principle be inferred by running them and performing cross-entropy benchmarking \cite{Boixo2018}, evaluating the cross-entropy requires exact simulation of the circuits in question and is infeasible except for small depth or qubit number. To estimate the expected state fidelity in RCS (and therefore the anticipated performance in cross-entropy benchmarking), we follow the strategy of Refs.~\cite{mayer2021theory, Proctor22, proctor2022measuring, morvan2023phase,DeCross2025} and infer the fidelity of a layer depth-$l$ circuit by computing the return-probability $F_{MB}$ of a ``mirrored'' layer depth-$l/2$ circuit, with the second (mirrored) half of the circuit employing randomized compiling to prevent unintended cancellation of coherent errors. The randomness for randomized compilation is sampled in real-time at the start of each shot, and the corresponding random 1Q gates are compiled on the fly (with the existing Haar-random 1Q gates), resulting in only one physical 1Q gate per qubit per layer. Following Ref.~\cite{DeCross2025}, we also initialize each mirrored circuit into a random computational basis state to prevent unequal SPAM errors between the two basis states from biasing the fidelity estimate. At each depth, we execute between 1000 and 2500 shots spread evenly across 100 random circuit connectivities. 

The fidelity of RCS as a function of depth inferred in this manner is reported in Fig.\,\ref{fig:rcs}a. We perform a least-squares best fit to the gate-counting model from \cite{DeCross2025}, 
\begin{align}
    F_{GC}(l) = (1 - p_{\text{spam}})^N (1 - \frac{5}{4} \epsilon_{\rm eff,2Q})^{\frac{N}{2} (l - \delta)}.
\end{align}
Here, $N = 98$, $\delta = 1.12$ is a correction to effective circuit layer depth from boundary effects in mirror circuits \cite{DeCross2025}, $p_{\text{spam}}$ is the effective SPAM error, and $\epsilon_{\rm eff,2Q}$ is the effective average 2Q error rate, which includes effects from 1Q, 2Q, and memory errors as in the previous section. From the fit, we estimate $p_{\text{spam}} = 5.3(51) \times 10^{-4}$ and $\epsilon_{\rm eff,2Q} = 2.00(6)\times 10^{-3}$. This effective 2Q error is also consistent with the estimate obtained from random Clifford circuits as well as component benchmarks reported in Table~\ref{tab:component benchmarks}.

Heuristic estimates of the classical cost of drawing samples from forward circuits at the same depths is shown in Fig.\,\ref{fig:rcs}b. The reported costs are for optimized tensor-network contraction assuming so-called ``embarrassing parallelization" (via slicing) into independent computations involving various amounts of available memory, and were obtained using (sliced) simulated annealing built into \texttt{cotengra} \cite{Gray2021hyperoptimized}. We note that the contraction-cost optimization performed here is only approximate, and the costs could certainly be mildly improved by providing the optimization heuristics with more computational power. However, we do not expect such improvements to change the overall conclusion that Helios can produce states at high global fidelity for which the (classical) sampling cost is vastly beyond the capabilities of existing supercomputers.

\section{Outlook}
\label{sec:outlook}

In this manuscript, we reported on how Helios operates and its current performance. Even at this early stage in its lifecycle, Helios exhibits state-of-the-art capabilities at the scale of $\sim100$ qubits. Like its predecessors Quantinuum H1 and H2, we expect Helios's performance to improve over time. Examples of relatively straight-forward performance improvements include: (1) fewer gate errors as our two-qubit gate error model suggests the 2Q gate error could be cut in half, (2) smaller memory errors using dynamic decoupling strategies~\cite{Biercuk2009} and (3) reduced circuit times from both faster transport operations~\cite{Bowler2012,Walther2012,Sterk2022} and better compilation methods.

Beyond these performance improvements, increasing clock speed is one scaling challenge for the QCCD platform. In this work we begin to address this issue through a fundamental architectural shift by parallelizing operations \cite{Brandl2017}. Previous generations, H1 and H2, used the same space for ground-state cooling and gating operations, with cooling operations being up to two orders of magnitude slower~\cite{Pino2020,moses2023race}. Helios, on the other hand, spreads the cooling operation over space to allow ions to spend less time in the zones used for 2Q gates. By increasing the ratio of cooling zones to gate zones, future QCCD-based QPUs can optimize the processor zone complexity while simultaneously increasing the clock speed.

While we do not yet fully understand the power or limitations of Helios, the combination of a new qubit choice, device architecture, and control software runtime already represents significant progress in the push for more powerful devices,  scalable architectures, and capabilities for fault-tolerant computation. Helios is far beyond the simulation abilities of classical computers, as evidenced by the RCS demonstration described above, and well poised to expand upon the set of tasks best suited for contemporary quantum computers. Indeed, as reported in Refs.\,\cite{fh_tbr, cr_tbr, pn_space_advantage}, Helios is already enabling advancements in quantum simulations of superconductivity and in cryptographic protocols to generate certified randomness. 

Looking further ahead, the successful integration of the four-way junction paves the way for much larger QCCD processors. Junction-based architectures should allow QCCD machines to maintain all-to-all connectivity for large numbers of qubits, opening the design space for fault-tolerance to high-efficiency encodings~\cite{Breuckmann2021}, transversal logic~\cite{RyanAnderson2022,RyanAnderson2024}, low-overhead magic state factories~\cite{Dasu2025}, and single-shot error correction~\cite{Campbell_2019,Berthusen2024,Cain2024}.

\section*{Acknowledgments}
We thank the entire Quantinuum team for numerous contributions that enabled this work.
%We specifically thank Jack Ross for editing the image in Fig.~\ref{fig:ions_image}.
The contributions of the SNL authors were funded in part by the U.S. Department of Energy, Office of Science, Office of Advanced Scientific Computing Research, Quantum Testbed Pathfinder Program
%. T.P. acknowledges support from 
and in part by 
an Office of Advanced Scientific Computing Research Early Career Award. Sandia National Laboratories is a multi-program laboratory managed and operated by National Technology and Engineering Solutions of Sandia, LLC., a wholly owned subsidiary of Honeywell International, Inc., for the U.S. Department of Energy's National Nuclear Security Administration under contract DE-NA-0003525. All statements of fact, opinion, or conclusions contained herein are those of the authors and should not be construed as representing the official views or policies of the U.S. Department of Energy or the U.S. Government.

\section*{Data availability}
Most of the component benchmarking data are available on the Quantinuum website and will be updated as Helios improvements are introduced.

\clearpage
\newpage

\section*{Appendices}
\beginappendix

\section{Hardware Details}
\subsection{Quantum logic}\label{sec: app quantum logic}

For 2Q gates, we create a M{\o}lmer-S{\o}rensen\cite{Sorensen2000} interaction by using pairs of Raman beams aligned at 90 degrees to each other with the $\delta \vec{k}$ aligned along the axes of BYYB crystals, and we use the axial stretch mode at 1.86 MHz to couple the internal states of the ions. The uncontrolled optical phase of the gate is removed using wrapper pulses to generate a $ZZ$ interaction~\cite{Lee05, Pino2020}. 2Q gates can be performed in four of the operation zones.

As in Ref.\,\cite{an2022high}, state-preparation uses narrow-band optical pumping by driving $S_{1/2}$ leakage states first to $D_{5/2}$ with a narrow linewidth 1762 nm laser, and then to $P_{3/2}$ $F=0$ using 614 nm light where it will decay back to $S_{1/2}$ leading to population accumulation in the qubit subspace. The measurement protocol begins by transferring the $\vert F, m_F \rangle = \vert 1,0 \rangle$ qubit state to the $D_{5/2}$ manifold (shelving) with the 1762 nm laser, and population remaining in the $S_{1/2}$ manifold is measured with resonant fluorescence. We reduce measurement crosstalk by shelving all qubits located in the quantum logic region before measurement if an entire batch of 16 qubits is to be measured, called ``protected measure." Furthermore, the end user can also choose to shelve both $\vert 1,0 \rangle$ and $\vert 2,0 \rangle$ qubit states and check for leakage out of the qubit subspace, called ``ternary measure", and then measure the qubit state by de-shelving one of the qubit states and applying resonant light to check for fluorescence. The ternary measurement doubles the measure time as shelving-and-detect needs to be performed twice. All shelving operations use multiple pulses (cabinet shelving) with different final states to exponentially reduce population transfer errors.

\subsection{Ground state cooling}

For the $^{171}$Yb$^+$ coolant ion, the nuclear spin $I=1/2$ allows for a fast frequency selective state-preparation scheme not reliant on a particular polarization \cite{PhysRevA.76.052314}. For ground state cooling, we use counter-propagating lin-perp-lin Raman beams aligned at 45 degrees to the crystal axis to get a $\delta \vec{k}$ projection on all three principal axes (the radial modes are rotated so as to not be orthogonal or parallel to the trap surface). The parallel cooling is achieved using 5 pairs of Raman beams detuned from the $S_{1/2}$ to $P_{1/2}$ transition near 369.4 nm. The laser beam angles are aligned to $45 \pm 0.2$ degrees with respect to the storage legs such that the beams can simultaneously intersect an operational zone in each leg. The beam waist focii are positioned between the two zones so each zone has the same beam waist and intensity. Carrier Rabi rates of up to 1 MHz are achieved in all zones for the cooling operations.  In this configuration, we perform sideband cooling sequences to achieve ground state cooling times of approximately 3 ms \cite{Pino2020}.

\subsection{Qubit frequency calibrations}

Helios operates with an externally imposed bias field of 3.95 G, making the qubit states approximate clock states, meaning they are naturally robust to magnetic field fluctuations with a second-order field coefficient of 488.8~Hz/G$^{2}$ (at zero field they become true clock-states that are insensitive to magnetic fields up to second order).

Variations in the qubit frequency arise primarily from the slow drift of magnetic fields at the level of $\sim200$~$\mu$G (over 24 hrs) and their gradient, as well as varying AC Zeeman shifts of the clock transition in $^{137}$Ba$^+$ from the trap RF current. To mitigate these effects, we employ a real-time spatial phase tracking routine~\cite{RyanAnderson2022}. The routine gets corrections to the reference qubit frequency from measurements of the average magnetic field in the quantum operation zones and the spatially-varying magnetic field in all 277 wells. After these calibrations, the routine applies the appropriate corrections to 1Q operations.

\section{Program profiling}
\label{sec:appendix_circuit_profiling}

To profile programs written in Guppy  (see Sec.~\ref{sec:compiler}), the compiled code is executed on a real-time control system simulator. Although this simulator is separate from Helios, it accurately captures timing information by using the same compiler, real-time software, and system settings used on Helios.

\begin{figure}[!t]
    \centering
    \includegraphics[width=1.0\linewidth]{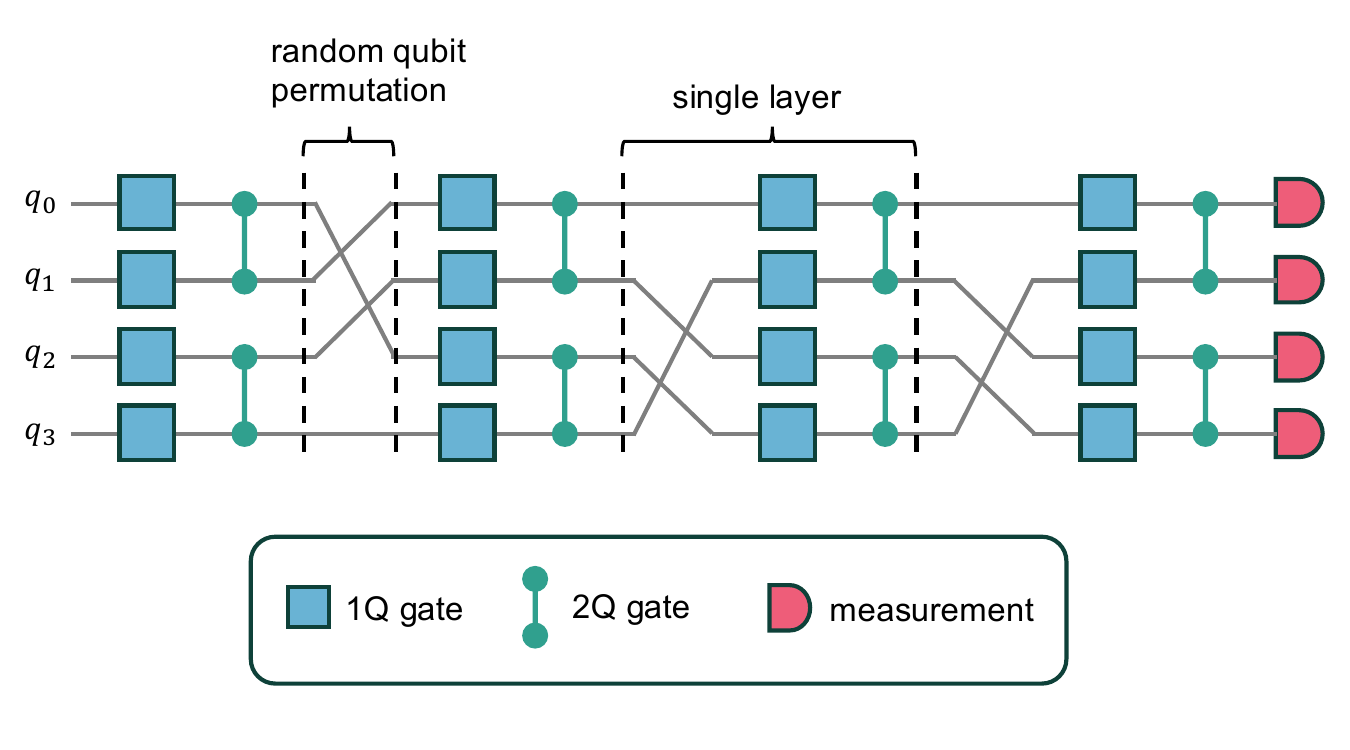}
    \caption{Example four-qubit program where all qubits are arbitrarily permuted and receive 1Q and 2Q gates each layer.}
    \label{fig:random_program_example}
\end{figure}

In Fig.~\ref{fig:depth1_time}b of the main text, we present timing results for three example programs run on the control system simulator. The first two programs consist of 1Q and 2Q gates executed on arbitrary qubit pairs, which are randomized each layer. Fig.~\ref{fig:random_program_example} illustrates their structure using an example four-qubit program with four layers. To get an accurate estimate of the time per layer, we exclude reset and measure operations that occur during the first and last layers of the program. We reduce the density of 2Q gates to roughly half by randomly selecting qubit pairs and gate them such that the 2Q gates occur $\left\lfloor d \frac{N}{2} \right\rfloor$ times per layer where $N$ is the number of active qubits in the program and $d=1$ ($d=\frac{1}{2}$) corresponds to fully (half) dense.

\begin{figure}[t]
    \centering
    \includegraphics[width=1.0\linewidth]{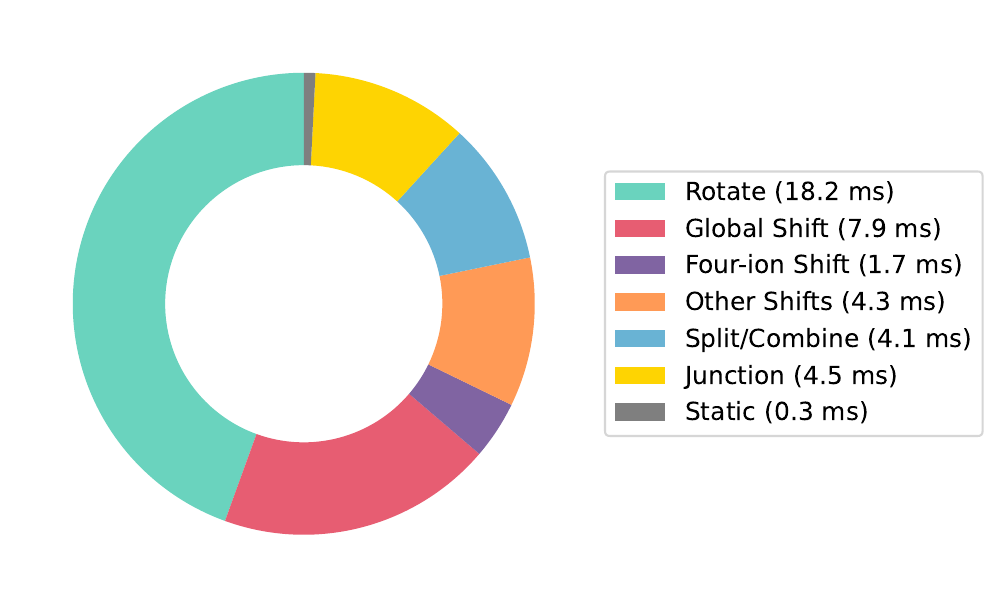}    \caption{Timing breakdown of transport operations for a single layer of the 98-qubit program profiled in Fig.~\ref{fig:depth1_time}a of the main text. Operations are broken down into the five categories described in Sec.~\ref{sec:TransportAndParallelOps}, plus four additional ones: global shift operations that collectively move ions in the cache, quantum operation, and storage leg regions; four-ion shift operations; all other shift operations; and static operations that do not move ions.}
    \label{fig:transport_timing}
\end{figure}

For the fully dense random program, Fig.~\ref{fig:transport_timing} shows a breakdown of the transport operation times per layer. Ring rotations dominate, while global shifts are the second-largest contributor. Future work will focus on reducing the total time spent on transport operations, thereby improving the depth-1 time. For example, compiler optimizations can reduce the number of transport operations in a program, while improvements in transport operation speed can lower their execution time.

The third program we profile uses 2D nearest-neighbor qubit pairing, which reflects common use cases such as quantum error correction and quantum simulation. While similar to the previously discussed example, this program restricts qubit pairings to one of four possible configurations on a square grid of qubits described as follows. Configurations (1) and (2) define horizontal pairings that alternate by row such that in (1), even rows pair adjacent qubits starting at the first column, while odd rows start at the second column; configuration (2) reverses this pattern. Configurations (3) and (4) follow the same alternating pattern but for vertical pairings, alternating the starting row by column parity: (3) starts at the first row for even columns and the second row for odd columns, while (4) reverses this pattern. Each layer in the program applies one of these pairing configurations, cycling through all four configurations every four layers.

\section{Benchmarking details}
\subsection{Correlated Error Analysis for randomized benchmarking} \label{app:correlation_analysis}
In this section, we detail our method for identifying correlated errors in simultaneous 1QRB and 2QRB experiments. Our analysis for correlations uses \emph{subsystem RB polarizations}. Consider an experiment of $k$ simultaneous RB experiments (e.g. qubits in operation zones of Helios). The result of running a simultaneous RB circuit consisting of parallel RB circuits on $k$ disjoint qubit subsets, labeled $1,2,\dots,k$, is described by a $k$-bit string $s_1s_2\dots s_k$, where $s_i = 0$ if the bit string outcome of the RB subexperiment on qubit subset $i$ matches its target outcome, and $s_i = 1$ otherwise. A \emph{subsystem parity} $z_S$ for $S \subseteq \{1, 2, \dots, k\}$ is defined as
\begin{align}
    z_S = \prod_{j \in S} (-1)^{s_j},
\end{align} 
and the expected value $\langle{z_S\rangle}$ is called a \emph{subsystem polarization}.

The \emph{subsystem decay factor} $\lambda_S$ is found by fitting the empirical values of $\langle z_S \rangle$ to an exponential decay, i.e. $\langle z_S \rangle_l = \lambda_S^l$, where $l$ is the circuit depth. In the absence of correlated errors, $\lambda_S = \prod_{k \in S}\lambda_k$. $\lambda_S$ is a Pauli channel eigenvalue for the twirled error channel produced by Clifford twirling on the individual qubit subsystems on which RB is performed in parallel. Each $\lambda_{\{j\}}$ is simply one of the eigenvalues of a Clifford-twirled error channel, and the total error rate of the channel is $\frac{4^n-1}{4^n}(1-\lambda_{\{j\}})$, where $n$ is the number of qubits in the RB subexperiment.

To search for evidence of correlated errors across multiple RB subexperiments, and quantify any such errors, we start by estimating $\lambda_{\{i\}}$ for each subexperiment $i$. Then, we estimate $\lambda_{\{i,j\}}$ for each pair $(i,j)$ of subexperiments and compute the test statistic 
\begin{align}
a = \log (\lambda_{\{i,j\}}) - \log(\lambda_{\{i\}}) - \log(\lambda_{\{j\}}).
\end{align}

In the absence of correlated errors between subsystems $i$ and $j$, $\lambda_{\{i,j\}} = \lambda_{\{i\}}\lambda_{\{j\}}$ and so $a=0$, and if there are any correlated error between subsystems $i$ and $j$, then  $\lambda_{\{i,j\}} > \lambda_{\{i\}}\lambda_{\{j\}}$ and so $a > 0$. In our analysis, we only have access to estimates of each $\lambda_{S}$, which we denote by $\tilde{\lambda}_S$, and a corresponding estimate of $a$, denoted $\tilde{a}$. We therefore compute $\tilde{a}$ and implement a statistical test to ascertain whether $\tilde{a}$ is large enough to conclude that $a > 0$ with at least 95\% confidence (an $\alpha = 0.05$ significance). We use a normal approximation for $\tilde{a}$'s distribution under the null hypothesis (no correlated errors), with a standard deviation for this distribution estimated using a non-parametric bootstrap. As we test for correlated errors between all $k$ choose $2$ pairs of subsystems, we implement our individual hypothesis tests (of whether each $a>0$) at $\frac{\alpha}{{k \choose 2}}= \frac{0.05}{{k \choose 2}}$ significance (a Bonferroni correction). This means that if there are no correlated errors, we will erroneously conclude there are correlated errors with at most 5\% probability, known as the family-wise error rate of the hypothesis tests.

We find that none of the $a$ are larger than zero, in our $95\%$ confidence hypothesis test, indicating no statistically significant evidence for correlated errors. Sufficiently small correlated errors will probably not be detected by this analysis. Using simulations, we can estimate what fraction of the error would have to be correlated error in order for our analysis to detect it. For the simultaneous 2QRB experiments we find that, given the estimated RB error rates from the experimental data, a two-subsystem correlated error would need to constitute approximately $10\%$ of the total error rate of the constituent subsystems (i.e., contributing $4.3 \times 10^{-4}$ to the average 2QRB infidelity) to be identified as statistically significant in our analysis with at least $50\%$ probability. For the 1QRB experiments, a correlated error would need to constitute approximately $50\%$ (i.e., contributing $1.19 \times 10^{-5}$ to the average 1QRB infidelity) of the total error rate of the constituent subsystems to be identified as statistically significant in our analysis with at least $50\%$ probability.

\subsection{Detailed component benchmarking data and experimental details}\label{sec: benchmarking data appendix}

For component-level benchmarks including 1QRB, transport-1QRB, 2QRB, CB, and MCMR crosstalk,
all error rates reported in the main text are the average infidelity, defined as follows.
Let $\mathcal{E}$ be the error process (a completely-positive map) for a given operation $\mathcal{U}$, such that its noisy implementation is given by $\mathcal{E}\circ\mathcal{U}$.
Then the average infidelity is
\begin{equation}\label{eq:average infidelity}
    \eps_{avg}(\mathcal{E}) = 1-\int d\psi\bra{\psi}\mathcal{E}(\ket{\psi}\bra{\psi})\ket{\psi},
\end{equation}
where the integral is taken over all pure states in the computational Hilbert space with respect to the Haar measure.

\subsubsection{SPAM}

For $a,b\in\{0,1\}$, let $p(a|b)$ denote
the probability of measuring outcome $a$ given state preparation $b$.
For the standard measurement,
we find $p(1|0)=8.1(1)\times 10^{-4}$ and $p(0|1)=1.6(5)\times10^{-4}$.
For the ternary measurement,
we find leakage probabilities of $p(L|0)=2.7(8)\times 10^{-3}$ and $p(L|1)=5.7(1)\times 10^{-3}$,
and SPAM errors of
$p(1|0)=7(1)\times 10^{-4}$ and $p(0|1)=2.8(2)\times 10^{-3}$,
conditioned on non-leakage.

\subsubsection{Single-qubit RB}

In 1Q Clifford RB, a sequence of $l$ uniformly random Clifford group elements are applied to a qubit,
followed by an inverse Clifford that randomly includes a bit-flip (X) gate.
In the absence of error, this process prepares the qubit in a random computational basis state.
In our decomposition of the 1Q Clifford group into native gates,
the 24 group elements have 0.375 pi/2 pulses and 0.75 pi pulses on average.

We run 1QRB simultaneously on 16 qubits in the 8 operation zones with different random sequences applied to each qubit~\cite{Gambetta2012}.
We use sequence lengths $l\in\{10, 1000, 2000\}$,
generate 10 circuits per each sequence length,
and run 100 shots of each circuit.
Table~\ref{tab: SQRB data} lists the measured leakage rates and average infidelities (including the contribution from leakage) for each individual qubit.

\begin{table}[H]
\caption{1QRB leakage rate and average infidelity.}
\label{tab: SQRB data}
\centering
{
\begin{tabular}{ccc}
\hline\hline
Qubit & Leakage Rate ($\times10^{-5}$) & Avg. Infidelity ($\times10^{-5}$) \\
\hline
$q_0$ & 0.9(1) & 2.9(8)\\
$q_1$ & 1.4(0) & 3.1(7)\\
$q_2$ & 1.4(1) & 2.3(4)\\
$q_3$ & 1.5(4) & 2.8(4)\\
$q_4$ & 1.2(3) & 2.4(4)\\
$q_5$ & 1.3(0) & 2.0(2)\\
$q_6$ & 0.8(2) & 2.7(4)\\
$q_7$ & 1.0(0) & 3.1(6)\\
$q_8$ & 1.5(2) & 2.5(3)\\
$q_9$ & 1.6(0) & 2.5(3)\\
$q_{10}$ & 0.9(5) & 2.1(3)\\
$q_{11}$ & 0.8(1) & 2.5(7)\\
$q_{12}$ & 1.2(1) & 2.3(3)\\
$q_{13}$ & 1.4(4) & 2.4(4)\\
$q_{14}$ & 0.5(2) & 2.1(4)\\
$q_{15}$ & 0.6(0) & 2.5(4)\\ 
\hline\hline
Mean & 1.12(7) & 2.5(1)\\ 
\hline\hline
\end{tabular}
}
\end{table}

\subsubsection{Two-qubit RB}

Like 1QRB, 2QRB is performed by executing
sequences of $l$ uniformly random Clifford group elements
(now drawn from the 2-qubit Clifford group). A final inverse Clifford
then ideally prepares the qubit pair in a random computational basis state.
The 2QRB circuits are performed on 8 pairs of qubits initialized in the 8 operation zones,
each with a distinct random sequence.
As described in Sec.~\ref{sec:TransportAndParallelOps}, the
2Q gates are applied in parallel in only four out of eight zones.
We select 8 pairs of qubits for benchmarking
as this configuration corresponds to a typical batch of parallel operations during circuit execution.

\begin{table}[H]
\centering
\caption{Two-qubit RB leakage rates and average infidelities. Following the protocol described in Sec.~\ref{sec:TransportAndParallelOps} for performing 2Q gates on eight qubit pairs in the operation zones using 2Q gates applied in only four zones, pairs $(0,1)$ and $(2,3)$ utilize the same zone for the 2Q gate operation, similarly for pair sets $(4,5)$, $(6,7)$ and $(8,9)$, $(10,11)$ and $(12,13)$, $(14,15)$. Most sets of the qubit pairs utilizing the same zone have infidelities and leakage rates that agree within uncertainties, to the extent there are differences they may arise from differences in the 1Q gates, memory errors, and cooling, which occur in the eight separate zones.}
\label{tab: TQRB data}
\begin{tabular}{ccc}
\hline\hline
Qubit Pair & Leakage Rate ($\times10^{-4}$) & Avg. Infidelity ($\times10^{-4}$) \\
\hline
$(0,1)$ & 1.8(3) & 6.1(5)\\
$(2,3)$ & 2.2(1) & 7.7(5)\\
$(4,5)$ & 2.6(2) & 8.2(6)\\
$(6,7)$ & 2.1(1) & 6.7(5)\\
$(8,9)$ & 2.7(1) & 8.0(5)\\
$(10,11)$ & 3.3(5) & 8.8(6)\\
$(12,13)$ & 2.3(2) & 8.7(5)\\
$(14,15)$ & 2.6(2) & 8.7(6)\\
\hline\hline
Mean & 2.4(1) & 7.9(2)\\ 
\hline\hline
\end{tabular}
\end{table}

\subsubsection{Two-qubit cycle benchmarking}

2QCB works by preparing eigenstates of a Pauli operator $P$,
applying a Pauli-twirled 2Q gate $l$ times, and measuring in the $P$ basis.
We Pauli-twirl~\cite{Wallman2016} the 2Q gates so that the error channel $\mathcal{E}$ can be assumed to be a stochastic Pauli channel,
which is defined as
\begin{equation}
    \mathcal{E}(\rho)=\sum_i p_i P_i\rho P_i,
\end{equation}
where the sum is over all Pauli operators modulo an overall phase,
and the $p_i$ are probabilities that sum to one.

The eigenoperators of any stochastic Pauli channel are themselves Pauli operators, so $\mathcal{E}(P_i)=f_i P_i$,
and their eigenvalues $f_i$ are often called Pauli fidelities and are given by
\begin{equation}
f_i = \sum_j (-1)^{\langle i,j\rangle}p_j,
\end{equation}
where the symbol $\langle i,j \rangle$ equals 0 or 1, depending on whether $P_i$ and $P_j$ commute or anti-commute, respectively.
The Pauli error probabilities can be computed from the Pauli fidelities via
\begin{equation}\label{eq: Hadamard transform}
    p_i = \frac{1}{d^2}\sum_j (-1)^{\langle i,j\rangle}f_j,
\end{equation}
where $d$ is the Hilbert space dimension.
Denote the noisy 2QCB circuit of length $l$ as $\mathcal{C}_l$.
2QCB estimates the Pauli fidelities by fitting the empirical expectation values $\mathbb{E}_l(P_i)=\mathrm{Tr}(P_i\mathcal{C}_l(P_i))$
to the model $\mathbb{E}_l(P_i)=Af_i^l$.
The Pauli error probabilities are then computed via Eq.~\eqref{eq: Hadamard transform}.
In terms of the Pauli error probabilities,
the average infidelity (not including leakage) is given by
\begin{equation}
\eps_{avg}(\mathcal{E})=\frac{d}{d+1}\sum_{i>0}p_i,
\end{equation}
where the sum is over all non-identity Pauli probabilities.

Because gate sets have a gauge freedom, not all $d^2$ individual Pauli fidelities can be learned in a SPAM robust way, but rather,
only the geometric means of subsets of Pauli fidelities that are related to each other by the action of the gate being benchmarked~\cite{Chen2023}.
We therefore assume that pairs of Pauli operators within the same orbit of $R_{ZZ}(\pi/2)$ have the same fidelity (for example: $f_{IX}=f_{ZY}$).
Furthermore,
simulations of known error sources in $R_{ZZ}(\pi/2)$ shows symmetry between $X$ and $Y$ in the
Pauli fidelities.
We therefore only estimate $f_i$
for $P_i\in\{IZ,ZI,ZZ,IX,XI,XX\}$,
and we assume any two Paulis that are related by an $X$-$Y$ symmetry to have the same fidelity (i.e., $f_{XY}=f_{XX}$).

In our 2QCB experiment,
we prepare the 8 states in the tensor product bases $\{\ket{0}, \ket{1}\}^{\otimes2}$ and $\{\ket{+}, \ket{-}\}^{\otimes2}$,
apply the $R_{ZZ}(\pi/2)$ gate $l$ times with $l\in\{4, 100, 200, 400\}$,
and measure each qubit in the same basis that it was prepared in.
For each state preparation and sequence length we run 200 shots and employ runtime randomness in the software stack to implement single-shot Pauli-twirling on the 2Q gates.
As in 1QRB and 2QRB,
we use the ternary measurement and fit the probability of not leaking versus $l$ to infer a leakage rate per gate.
The non-leaked population is then used to compute expectation values that decay with $l$.
We perform the experiment in parallel on 8 qubit pairs initialized in the 8 operation zones as in 2QRB and randomize the order of state preparations within each zone.
The leakage rates and average infidelities (including leakage) are listed in Tab.~\ref{tab: TQCB data}.
The zone-averaged Pauli error probabilities,
up to unlearnable degrees of freedom and symmetry assumptions,
are listed in Tab.~\ref{tab: TQCB data2}.

\begin{table}[H]
\caption{2QCB estimated leakage rates and infidelities. Qubit pairs that share zones are the same as in \ref{tab: TQRB data}}
\label{tab: TQCB data}
\centering
{
\begin{tabular}{ccc}
\hline\hline
Qubit Pair & Leakage Rate ($\times10^{-4}$) & Avg. Infidelity ($\times10^{-4}$) \\
\hline
$(0,1)$ & 1.0(1) & 9.6(4)\\
$(2,3)$ & 1.0(1) & 9.6(5)\\
$(4,5)$ & 1.0(1) & 6.1(4)\\
$(6,7)$ & 1.0(2) & 6.0(4)\\
$(8,9)$ & 1.0(2) & 5.9(3)\\
$(10,11)$ & 1.1(1) & 7.4(4)\\
$(12,13)$ & 1.6(1) & 10.1(4)\\
$(14,15)$ & 1.6(1) & 9.7(5)\\
\hline\hline
Mean & 1.14(6) & 8.1(2)\\ 
\hline\hline
\end{tabular}
}
\end{table}

\begin{table}[!h]
\caption{2QCB estimated Pauli error probabilities,
averaged over all qubit pairs,
up to unlearnable degrees of freedom and symmetry assumptions.
For error classes with greater than one element,
the right column is the probability of every Pauli error in the set.}
\label{tab: TQCB data2}
\centering
{
\begin{tabular}{cc}
\hline\hline
Error Class & Probability ($\times10^{-5}$)\\
\hline
$\{IX,IY,ZX,ZY\}$ & 4.5(2)\\
$\{XI,YI,XZ,YZ\}$ & 5.8(2)\\
$\{XX,XY,YX,YY\}$ & 0.06(4)\\
$\{IZ\}$ & 19(1)\\
$\{ZI\}$ & 19(1)\\
$\{ZZ\}$ & 5.9(9)\\
\hline\hline
\end{tabular}
}
\end{table}

\subsubsection{Transport-1QRB}
We ran transport-1QRB with $k\in\{1, 2, 4, 8\}$ and sequence lengths $l\in\{8, 64, 128\}$,
where sequence length here refers to the number of depth-1 transport operations. 
For each sequence length, we generate 10 circuits and run each circuit for 100 shots. 
\begin{table}[H]
\centering
\caption{Transport-1QRB leakage rates and average infidelities.
Transport depth is the number of depth-1 transport operations between 1Q Cliffords.
The reported numbers are averaged over qubits with a given transport depth.}
\label{tab: transport SQRB data}
\resizebox{\columnwidth}{!}
{
\begin{tabular}{ccc}
\hline\hline
Transport Depth & Avg. Leakage Rate ($\times10^{-4}$) & Avg. Infidelity ($\times10^{-4}$) \\
\hline
1 & 4.4(2) & 6.0(3)\\
2 & 8.3(5) & 12.8(7)\\
4 & 17.1(5) & 34(2)\\
8 & 35(2) & 84(5)\\
\hline\hline
\end{tabular}
}
\end{table}

\subsubsection{MCMR crosstalk test} \label{app:mcmr_errors}

We quantify MCMR crosstalk errors by fitting the spectator qubit survival probabilities to a linear decay model as a function of the number of applied MCMRs to the target qubits. We relate the fit parameters to error magnitude based on an effective quantum jump operator description of the error channel. This is an expansion of previous work on ``bright-state depumping" for $^{171}$Yb$^+$ qubits~\cite{Gaebler2021crosstalk} where the decay rate of spectator qubits prepared in the $\ket{1}$ state was used to determine the average infidelity. For $^{137}$Ba$^+$ qubits, however, the added complexity of the crosstalk decay channels requires that the spectator qubits be prepared in additional states.  Furthermore, the ternary measurement is used to resolve bit-flip from leakage errors.  

The MCMR crosstalk error channel is modeled as a set of effective quantum jump operators $\hat{L}_{ij}=\sqrt{\gamma_{ij}}|i\rangle\langle j|$ occurring at rates $\gamma_{ij}$ between states $i$ and $j$ with $i,j\in \{0, 1, L\}$, leading to population transfer and decoherence~\cite{Molmer:93,PhysRevA.85.032111}. Evaluating Eq.~\eqref{eq:average infidelity} results in an average infidelity
\begin{align}\label{eq:crosstalkec}
    \epsilon_{avg}(\mathcal{E})
    =& \frac{1}{6}\left(p(0|1)+p(1|0)+2p(L|0) +2 p(L|1) + 4p_Z\right)
\end{align}
where $\mathcal{E}$ is the crosstalk error channel, $\epsilon_{avg}(\mathcal{E})$ is the average infidelity, $p(i|j)$ is the conditional probability for transitioning from state $j$ to state $i$ via a quantum jump, and $p_Z$ is the phase-flip probability.  Individual terms in Eq.~\eqref{eq:crosstalkec} can be resolved by preparing spectator qubits in eigenstates of the Pauli operators~\cite{BOWDREY2002258} and using the ternary measurement.  In Sec.~\ref{sec:mcmr}, circuits preparing the spectator qubits in each state of the computational basis were used to estimate bit-flip and leakage probabilities with results shown in Fig.~\ref{fig: mcmr crosstalk}.  

Measuring $p_Z$ requires circuits preparing the spectator qubits in the X/Y eigenstates, which suffer from additional memory error that we separately quantify with transport-1QRB (see Sec.~\ref{sec:memory_error}).  Instead, to estimate $p_Z$ (and consequently $\epsilon_{avg}(\mathcal{E})$), we expand $p_Z\approx [p(1|0)+p(0|1)+p(L|1)+p(L|0) + p_{el}]/4$, which reflects the scattering-induced random phase shifts leading to crosstalk-induced decoherence.  This expansion makes an assumption that the intensity of crosstalk light is weak such that the duration of an MCMR on a target qubit is brief compared to the crosstalk transition rates $\gamma_{ij}$, which is well-satisfied in practice.  The elastic (Rayleigh) contribution $p_{el}$ was measured in Ref.~\cite{PhysRevLett.105.200401} using a spin-echo sequence for $^{9}\text{Be}^{+}$.  We estimate the contribution of $p_e$ to the average infidelity to be roughly $\% 8$ of the total error budget, however measurement of this contribution on Helios remains the subject of future study.  
Measuring $p_Z$ requires circuits preparing the spectator qubits in the X/Y eigenstates, which suffer from additional memory error that we separately quantify with transport-1QRB (see Sec.~\ref{sec:memory_error}).  Instead, to estimate $p_Z$ (and consequently $\epsilon_{avg}(\mathcal{E})$), we expand $p_Z\approx [p(1|0)+p(0|1)+p(L|1)+p(L|0) + p_{el}]/4$, which reflects the scattering-induced random phase shifts leading to crosstalk-induced decoherence.  This expansion makes an assumption that the intensity of crosstalk light is weak such that the duration of an MCMR on a target qubit is brief compared to the crosstalk transition rates $\gamma_{ij}$, which is well-satisfied in practice.  The elastic (Rayleigh) contribution $p_{el}$ was measured in Ref.~\cite{PhysRevLett.105.200401} using a spin-echo sequence for $^{9}\text{Be}^{+}$.  We estimate the contribution of $p_e$ to the average infidelity to be roughly $8\%$ of the total error budget, however measurement of this contribution on Helios remains the subject of future study.  

\begin{figure}[t]\label{fig: crosstalk all-even}
\includegraphics[width=\columnwidth]{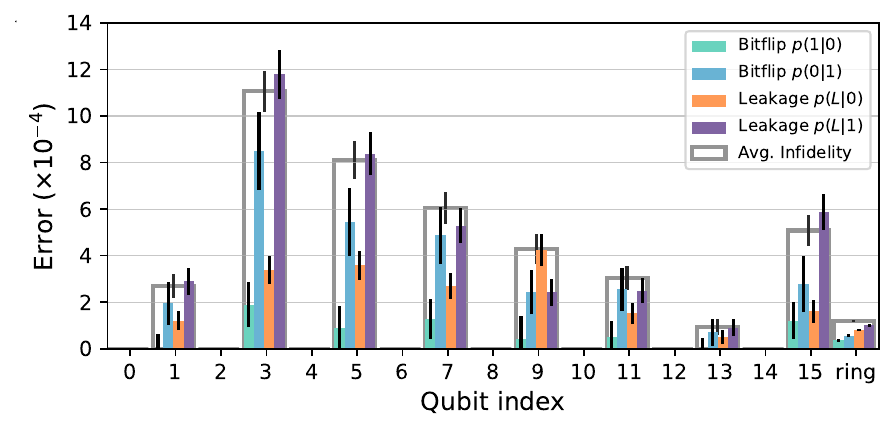}
\caption{MCMR crosstalk data with target qubits in \{0, 2, 4, 6, 8, 10, 12, 14\}.
MCMR operations are applied simultaneously to the target qubits.
The plot shows estimated rates of different error channels (smaller bars) and the average infidelity (wider bars) for spectator qubits in the operation zones (qubit index in \{1, 3, 5, 7, 9, 11, 13, 15\}) and in the storage ring.
}
\end{figure}

In addition to the MCMR crosstalk experiments described in Sec.~\ref{sec:mcmr},
we also run an MCMR crosstalk experiment
on 8 target qubits simultaneously.
This arrangement has one target qubit and one spectator qubit in each operation zone,
with the remaining qubits in the storage ring as spectator qubits.
We perform this experiment to estimate the contribution of MCMR crosstalk to the effective MCMR error in the system-level random Clifford circuits benchmark,
for circuits that contain batches of multiple MCMRs per layer (see Sec.~\ref{sec: binary RB}).
The data is shown in Fig.~\ref{fig: crosstalk all-even}.
We find an average MCMR crosstalk error per qubit of $5.2(2)\times10^{-5}$ in the operation zones and $1.21(4)\times10^{-5}$ in the storage ring.

\begin{table}[H]
\centering
\caption{Local and global MCMR crosstalk error channels estimated from Fig.~\ref{fig: mcmr crosstalk}.}
\label{tab: MCMR crosstalk data}
\begin{tabular}{ccc}
\hline\hline
Error Channel & Local ($\times10^{-4}$) & Global ($\times10^{-5}$) \\
\hline
$p(1|0)$ & 0.7(2) & 1.2(2)\\
$p(0|1)$ & 1.6(2) & 2.8(2)\\
$p(L|0)$ & 0.8(1) & 2.1(2)\\
$p(L|1)$ & 1.8(2) & 4.8(2)\\
\hline\hline
\end{tabular}
\end{table}

\subsection{Random Clifford circuits with mid-circuit measurements}\label{sec: app: QiRB}

Here we provide additional details on the system-level random Clifford circuits with MCMR benchmark,
discussed in Sec.~\ref{sec: binary RB}.
We first describe the method of stabilizer tracking.

A stabilizer of a state $\ket{\psi}$ is a Pauli operator for which
$\ket{\psi}$ is a +1 eigenstate.
Initially, the state is $\ket{0\dots0}$ and is stabilized by all Paulis tensor products of $I$ and $Z$.
We select a random initial stabilizer $S$
with biased sampling of $\{I,Z\}$ with probabilities $\{1/4, 3/4\}$.
We then propagate $S$ through each layer of the circuit
using an efficient binary matrix representation of the Clifford operations.
When a qubit is chosen for an MCMR operation, a 1Q correction gate is applied to measure the qubit in the Pauli basis determined by $S$ (or equivalently, since all measurements are in the $Z$ basis, the correction maps the stabilizer from $X$ or $Y$ to $Z$).
After an MCMR, a new qubit is appended to $S$ with stabilizer randomly randomly chosen from $\{I,Z\}$ with the same biased sampling of probabilities in $\{1/4, 3/4\}$.
Finally, at the end of the circuit, each qubit is measured in the Pauli basis according to $S$.
The shot succeeds if the parity of the measured bitstring (including all the mid-circuit measurements) agrees with the sign of the evolved stabilizer $S$.

In our experiment we choose $n_m\in\{0,8,16\}$ and $l\in\{2,4,6,8\}$.
For each value of $n_m$ and $l$,
we generate 10 circuits,
and run each circuit for 100 shots,
with the order of all circuits randomized.
The average success probability is rescaled into a quantity called the polarization~\cite{Hines2024}, defined as $y_{pol}=2p_{succ}-1$.
Let $F(n_m)$ be the process fidelity per circuit layer
as a function of $n_m$.
We estimate $F(n_m)$ by fitting the polarization to the model $y_{pol}(l, n_m)=AF(n_m)^l$,
where $A$ is a 98-qubit SPAM parameter that we fix to be equal for all values of $n_m$.

Below we list the process fidelities per circuit layer for $n_m\in\{0,8,16\}$,
which are plotted in Fig.~\ref{fig:binary RB fig} in Sec.~\ref{sec: binary RB}.
Here $n_m$ is the number of MCMRs per circuit layer.

\begin{table}[H]
\caption{Process fidelity per layer versus number of mid-circuit measurements and resets per layer.}
\label{tab: binary RB layer fidelities}
\centering

{
\begin{tabular}{ccc}
\hline\hline
MCMRs per layer & Layer fidelity\\
\hline
0 & 0.883(16)\\
8 & 0.856(15)\\
16 & 0.862(15)\\
\hline\hline
\end{tabular}
}
\end{table}

As explained in Sec.~\ref{sec: binary RB},
we compute an effective 2Q gate error $\epsilon_{\mathrm{eff},2Q}$ from the $n_m=0$ data,
and separate effective MCMR errors $\epsilon_M$ from both the $n_m=8$ and $n_m=16$ data.
The effective fidelities as well their predicted values from the component-level benchmarking data are listed in Tab.~\ref{tab: binary RB effective fidelities}.
Here, we explain our procedure for predicting the effective fidelities.

For $\epsilon_{\mathrm{eff},2Q}$, we take the 2Q gate error from 2QRB in Tab.~\ref{tab:component benchmarks},
and we convert the average infidelity into a process infidelity.
We then take the depth-1 memory error per qubit from Tab.~\ref{tab: transport SQRB data},
again convert into a process infidelity and multiply by two (to get depth-1 error per two qubits).
We add the process infidelities from the 2Q gate to the memory error and convert again into average infidelity:
\begin{equation}
\epsilon_{\mathrm{eff},2Q}=\frac{4}{5}\bigg(\bigg(\frac{5}{4}\bigg)\eps_{avg,2Q}+2\bigg(\frac{3}{2}\bigg)\eps_{\mathrm{mem}}\bigg).
\end{equation}
Plugging in $\eps_{avg,2Q}=7.9(2)\times10^{-4}$ and $\eps_{\mathrm{mem}}=6.0(3)\times10^{-4}$ and propagating uncertainties gives the value of $\eps_{\mathrm{eff},2Q}$ in Tab.~\ref{tab: binary RB effective fidelities}.

For $\epsilon_M(n_m=8)$, we take the SPAM error of the standard measurement from Tab.~\ref{tab:component benchmarks},
and we add the measured crosstalk error from the MCMR crosstalk experiment with 8 simultaneous target qubits shown in Fig.~\ref{fig: crosstalk all-even},
since that is the measurement configuration used for the MCMRs in the $n_m=8$ QiRB circuits.
We add the measured crosstalk error per MCMR
in the storage ring times the number of spectator qubits in the ring,
plus the crosstalk error per MCMR in the operation zones times the number of spectator qubits in the zones:
\begin{multline}
\eps_M(n_m=8)=\eps_{SPAM} + 82\times\eps_{MCMR,\, \mathrm{ring}}\\
+ 8\times\eps_{MCMR,\, \mathrm{zones}}.
\end{multline}

For $\eps_M(n_m=16)$, since the batch of 16 measurements is performed using the ``protected measure" scheme,
we omit the crosstalk error on qubits in the operation zones:
\begin{equation}
\eps_M(n_m=16)=\eps_{SPAM} + 82\times\eps_{MCMR,\, \mathrm{ring}}.
\end{equation}
Plugging in with $\eps_{SPAM}=4.8(6)\times10^{-4}$, $\eps_{MCMR,\,\mathrm{ring}}=1.51(5)\times10^{-5}$, and $\eps_{MCMR,\,\mathrm{zones}}=6.5(3)\times10^{-5}$,
and propagating uncertainties gives the values of $\eps_M$ listed in Tab.~\ref{tab: binary RB effective fidelities}.

\begin{table}[!h]
\caption{Effective fidelities estimated from the random Clifford with MCMR circuits data (middle column),
compared to predicted values from the component-level benchmarking data (right column).}
\label{tab: binary RB effective fidelities}
\centering
{
\begin{tabular}{ccc}
\hline\hline
Parameter &  System-level & Component-level\\
& value ($\times10^{-3}$)& value ($\times10^{-3}$)\\
\hline
$\epsilon_{\mathrm{eff},2Q}$ & 2.0(3) & 2.2(1)\\
$\epsilon_M(n_m=8)$ & 2.6(13) & 2.2(1)\\
$\epsilon_M(n_m=16)$ & 1.0(7)& 1.7(1)\\
\hline\hline
\end{tabular}
}
\end{table}

\clearpage

\bibliography{Helios_truncated}

\end{document}